\documentclass[12pt]{amsart}
\usepackage{amsfonts}
\usepackage{amssymb}
\usepackage[dvips]{graphics}
\usepackage{epsfig}
\usepackage{euscript}
\usepackage{color}
\newcommand{\Rl}{\mathbb{R}}
\newcommand{\Nl}{\mathbb{N}}
\newcommand{\Ir}{\mathbb{Z}}
\newcommand{\Cx}{\mathbb{C}}

\newtheorem{theorem}{Theorem}[section]
          
\newtheorem{lemma}[theorem]{Lemma}              
              
\newtheorem{proposition}[theorem]{Proposition}
\newtheorem{corollary}[theorem]{Corollary}
\newtheorem{definition}[theorem]{Definition}
\newtheorem{conjecture}[theorem]{Conjecture}
\newcommand{\rem}[1]{{\bf Remark:}}

\newcommand{\Section}[1]{\setcounter{equation}{0}\section{#1}}

\newcommand{\eq}[1]{(\ref{#1})}
\renewenvironment{proof}{\noindent {\bf Proof: }}{\QED\medskip}
\def\QED{{\hspace*{\fill}{\vrule height 1ex width 1ex }\quad} 
    \vskip 0pt plus20pt}
\newcommand{\be}{\begin{equation}}
\newcommand{\ee}{\end{equation}}
\newcommand{\bea}{\begin{eqnarray}}
\newcommand{\eea}{\end{eqnarray}}
\newcommand{\beann}{\begin{eqnarray*}}
\newcommand{\eeann}{\end{eqnarray*}}

\newcommand{\ket}[1]{\vert{#1}\rangle}

\newcommand{\unity}{{1\hskip -3pt \rm{I}}}
\newcommand{\GS}{\mathcal{G}}
\newcommand{\ip}[2]{\langle{#1}|{#2}\rangle}
\newcommand{\Hil}{\mathcal{H}}

\newcommand{\Spin}{{\textrm{J}}}

\newcommand{\Obs}{\mathcal{A}}

\newcommand{\floor}[1]{\left\lfloor{#1}\right\rfloor}
\newcommand{\ceil}[1]{\left\lceil{#1}\right\rceil}
\newcommand{\sech}{\, {\rm sech}}

\newcommand{\scrH}{\EuScript{H}}

\begin{document}

\begin{center}
{\LARGE \bf The Spectral Gap for the Ferromagnetic Spin-$\Spin$ XXZ Chain \\[27pt]}
{\large \bf Tohru Koma$^1$, Bruno Nachtergaele$^2$,
and Shannon~Starr$^3$\\[10pt]}
\end{center}
$^1$Department of Physics,
Gakushuin University,
Mejiro, Toshima-ku, Tokyo 171, Japan,
{tohru.koma@gakushuin.ac.jp}\\[10pt]
$^2$Department of Mathematics,
University of California, Davis,
Davis, CA 95616-8633, USA,
{bxn@math.ucdavis.edu}\\[10pt]
$^3$Department of Physics,
Princeton University,
Princeton, NJ 08544, USA,
{sstarr@math.princeton.edu}\\[10pt]

\noindent
{\bf Abstract}

We investigate the spectrum above the kink ground states
of the spin $\Spin$ ferromagnetic XXZ chain with Ising anisotropy
$\Delta$. Our main theorem
is that there is a non-vanishing gap above all ground states
of this model for all values of $\Spin$. Using a variety of methods,
we obtain additional information about the magnitude of this gap,
about its behavior for large $\Delta$, about its overall behavior as a
function of $\Delta$ and its dependence on the ground state, about the
scaling of the gap and the structure of the low-lying spectrum for large
$\Spin$, and about the existence of isolated eigenvalues in the excitation
spectrum. By combining information obtained by perturbation theory,
numerical, and asymptotic analysis we arrive at a number of interesting
conjectures. The proof of the main theorem, as well as some of the
numerical results, rely on a comparison result with a Solid-on-Solid
(SOS) approximation. This SOS model itself raises interesting questions in
combinatorics, and we believe it will prove useful in the study of
interfaces in the XXZ model in higher dimensions.

\vspace{8pt}
\newline
{\small \bf MCS numbers:} 82B10, 82B24, 82D40 
\vfill
\hrule width2truein \smallskip {\baselineskip=10pt \noindent Copyright
\copyright\ 2001 by the authors. Reproduction of this article in its entirety, 
by any means, is permitted for non-commercial purposes.\par }

\newpage

\section{Introduction}
The subject of our paper is the ferromagnetic XXZ model.
The XXZ model is one of the best studied quantum spin systems,
benefiting from both algebraic and analytic techniques.
But, while the mathematical techniques that have been applied to the XXZ model 
are impressive (c.f.\ in \cite{JM}), 
many of the most basic physical questions remain open \cite{MN}.
We address an unresolved issue of the one-dimensional XXZ model, which is the following.
It is known, by rigorous methods \cite{KN1}, that there is a spectral gap above
the infinite volume ground state for spin $1/2$,
but the proof relies on an algebraic tool which is not present for higher spins.
How does one prove the existence of a spectral gap in the more general setting?
We answer the question in the present paper.
Our methods are somewhat more general than those of \cite{KN1}
since we do not rely on the quantum group symmetry.
On the other hand, it is essential for our proof that we know the spectral
gap exists for the spin $1/2$ XXZ model.
Still, we believe our techniques may be applied to other spin models,
as well as shedding light on this corner of the general knowledge of the XXZ model.

The XXZ spin chain is a generalization of the Heisenberg model where one allows 
anisotropic spin couplings.
The Hamiltonian for the spin $\Spin$ model is
\begin{equation}
\label{intro hamiltonian}
H^{(\Spin)}_\Lambda = - \sum_{\langle{\alpha,\beta}\rangle \in \Lambda} (S_\alpha^1 S_{\beta}^1 + S_\alpha^2 S_\beta^2
 + \Delta S_\alpha^3 S_\beta^3)\, ,
\end{equation}
where $S_\alpha^{1,2,3}$ are the spin $\Spin$ matrices acting on the site $\alpha$, tensored with the identity
operator acting on the other sites. $\langle{\alpha,\beta}\rangle$ denotes
a pair of nearest neighbors.
The local Hilbert space is $\Hil_\alpha \cong \Cx^{2\Spin+1}$, and $H_\Lambda$ is a Hermitian operator
on $\Hil_\Lambda = \bigotimes_{\alpha \in \Lambda} \Hil_\alpha$.
For now we think of $\Lambda$ as a finite subset of $\Ir$, though we are also interested in the
case that $\Lambda = \Nl$ or $\Ir$.
The main parameter of the model is the anisotropy $\Delta \in \Rl$.
By choosing $\Delta = \pm 1$ we can obtain the isotropic ferromagnet or antiferromagnet.
Alternatively, by taking $\Delta \to \pm \infty$ we recover the spin $\Spin$ Ising ferromagnet
and antiferromagnet.
In this paper we restrict $\Delta > 1$, which corresponds to a ferromagnet with the
strongest coupling along the $S^3$-axis.
We note that $H_\Lambda(\Delta)$ and 
$-H_\Lambda(-\Delta)$ are unitarily equivalent.
It is useful to introduce two other forms of $\Delta$:
$q = \Delta - \sqrt{\Delta^2 - 1}$ and $\eta = - \log q$.
Observe that $\Delta = \frac{1}{2}(q+q^{-1}) = \cosh(\eta)$.
The parameter $q$ is the one which labels the quantum group $\textrm{SU}_q(2)$ when $\Spin=1/2$.
As $\Delta$ increases from $1$ to $\infty$, $q$ decreases from $1$ to $0$, and $\eta$ 
increases from $0$ to $\infty$.

For $\Delta > 1$, it is widely known that
there are two infinite volume ground states which correspond to all spins up, $\ket{+\Spin}$
and all spins down, $\ket{-\Spin}$. 
It is considerably less well known that there are in fact
many more infinite volume ground states. These extra ground states come in two
families: the kink states and the antikink states. The kink states are an
infinite family of ground states all with the same GNS space, which break
discrete translation symmetry as well as the continuous $\textrm{U}(1)$
symmetry associated to the XXZ model. They have the property of being
asymptotically all down spins at $-\infty$ and all up spins at $+\infty$. They
clearly also break left-right symmetry: their reflected counterparts are the
antikink states. 
Our main results concern the spectral gap above these kink ground states.
The kink states are physically
interesting for several reasons: They exhibit domain walls, an
important feature of real ferromagnets (see \cite{NS} for an application of the
XXZ model to spin droplets in one dimension); For $\Spin>1$ the kink
ground states of the XXZ model are more stable than the Ising ground states, 
which is a new result and subject of the present paper.
Also, the XXZ spin chain plays an important role in explaining the phenomenon
of negative resistance jumps and hysteresis in recent magnetoresistance experiments
\cite{KY1, KY2}.

Although our main subject is the XXZ model for $\Spin > 1/2$,
let us briefly recall some important facts about the spin $1/2$ model. 
What is probably most well known is that the spin $1/2$ model is Bethe 
ansatz solvable.
This is not applicable to our case however, and we will not use the Bethe
ansatz in any way.
A second interesting feature of the spin $1/2$ XXZ model is 
that it possesses a quantum group symmetry.
Specifically, in \cite{PS} it was shown that adding a boundary field
\begin{equation}
\label{boundary field}
B = \Spin \sqrt{\Delta^2 - 1} (S_1^3 - S_L^3)
\end{equation}
makes the Hamiltonian commute with $\textrm{SU}_q(2)$ on $\Hil_\Lambda$.
(Actually there are two representation of $\textrm{SU}_q(2)$
corresponding to the two opposite linear orderings of its tensor factors;
$H_\Lambda + B_\Lambda$ commutes with one representation
and $H_\Lambda - B_\Lambda$ commutes with the other.
We will only consider $H_\Lambda + B_\Lambda$.)
For $0<q<1$ the representation theory of $\textrm{SU}_q(2)$ is 
equivalent to that of $\textrm{SU}(2)$ (c.f. \cite{Kas}),
and it plays the same role in the analysis of the XXZ model that $\textrm{SU}(2)$
plays in the analysis of the isotropic model. 
For example, the ground state space corresponds to the 
highest-dimensional irreducible representation of $\textrm{SU}_q(2)$.
In \cite{KN1}, Koma and Nachtergaele used the quantum group symmetry to 
calculate the spectral gap for the XXZ model by proving that the lowest excitations
of the XXZ model form a next-highest dimensional irreducible representation of
$\textrm{SU}_q(2)$.
There are still open conjectures relating to the representations of 
$\textrm{SU}_q(2)$ and the XXZ model such as: Prove the lowest (highest) energy of $H_\Lambda + B_\Lambda$
restricted to the spin $s$ representations in $\Hil_\Lambda$ is lower than the lowest (highest)
energy of the spin $s-1$ representations.
This is almost certainly true, and would generalize the Lieb and Mattis result \cite{LM},
but remains open.

We now turn our attention to the spin $>1/2$ models. The first important fact
is that the ground states have been explicitly calculated for all finite
volumes, in all dimensions, all choices of spin, and even allowing different
values of anisotropy along bonds in the different coordinate directions
\cite{ASW}. In \cite{GW} the ground states were independently dicovered for
$\Spin=1/2$ and one dimension, and these ground states were generalized to
infinite volume ground states. They have the property of being frustration
free, which means that they not only minimize the expectation of the infinite
volume Hamiltonian, they minimize every nearest neighbor interaction, too.
Gottstein and Werner found all the frustration free ground states, and
conjectured that there were no other ground states. In \cite{Mat}, their
conjecture was proved correct, and the analogous statement for $\Spin>1/2$ was
proved in \cite{KN3}. Thus, one has a complete list of ground states for the one
dimensional XXZ model with any choice of $\Spin$. Unfortunately, the results on
infinite volume ground states are valid only in one dimension. Finding the
complete set of ground states in dimensions two and higher is an important open
problem.

In this paper we prove that there is a nonvanishing spectral gap above
the infinite volume ground states for every $\Spin$, thus extending the results
of \cite{KN1}.
We mention that the existence of a spectral gap is generally believed to follow from the
fact that the quantum interface of the kink ground states is exponentially localized.
Our results verify the conventional wisdom, and our proof does rely on the exponentially
localized interface.
However there are other important elements to our proof:
most notably, a rigorous comparison of the spin $\Spin$ chain with a 
spin $1/2$ ladder with $2\Spin$ legs.
The spin $\Spin$ XXZ chain is a quantum many body Hamiltonian.
The Hilbert space for the $L$-site spin chain is $(2\Spin+1)^L$, its dimension
grows exponentially with $L$. 
The dimension of the spin ladder is even larger, at $2^{2 \Spin L}$.
But an important bound, Lemma \ref{Key lemma}, allows us to restrict attention
to an $(L+1)^{2\Spin}$ dimensional subspace.
This allows the proof of the existence of the spectral gap, and also allows 
more efficient numerical methods for studying the XXZ model.
The reduced system resembles a quantum solid on solid model for the spin ladder.
We view the present problem as a warm up for the QSOS method, which we believe 
will play an important role in proving stability of the 111 interface for the 
XXZ model.
We also expect the spin ladder technique will be useful in proving the existence of a spectral
gap for other spin chains where a gap is known in $\Spin=1/2$ but not for $\Spin > 1/2$.
In this paper, in addition to giving a rigorous proof of the existence of a gap, we present
a new type of numerical method for studying the XXZ model.
We also present an asymptotic model for the low lying spectrum of the XXZ model
as $\Spin \to \infty$, in terms of a free Bose gas.
The asymptotics explain new qualitative features of the XXZ model for $\Spin>1/2$,
and is in excellent agreement with numerical data for $\Spin$ sufficiently high.

The remainder of the paper is organized as follows.
In Section 2 we present our main theorem, as well as a number of conjectures which 
are supported by numerical evidence and asymptotic analysis.
In Section 3 we introduce some background material which is useful for our proof.
In Section 4 we derive the spin chain / spin ladder reduction.
In Section 5 we finish the proof of the main theorem.
In Section 6 we combine the lower bounds for the spectral gap with numerical methods 
to obtain data for the spectral gap.
In Section 7 we derive a boson model for the XXZ spin system which explains
the asymptotic behavior of the gap as $\Spin \to \infty$.
This boson model is similar to \cite{HP,Dys1,Dys2}, but without the need
for a large external field (other than the boundary field which vanishes in
the thermodynamic limit).


\section{Main Result and Conjectures}
(The notation $[a,b]$ will always
refer to the discrete interval $\{a,a+1,\dots,b\}$.
It is not necessary that $a$ and $b$ are integers as long
as the difference $b-a$ is.)

The Hamiltonian we will use is the following spin $\Spin$
XXZ Hamiltonian:
\begin{equation}
\label{Hamiltonian definition}
\begin{split}
H^{\Spin}_\Lambda &= \sum_{\{\alpha,\alpha+1\} \subset \Lambda} 
h^\Spin (\alpha,\alpha+1) \\
h^\Spin (\alpha,\alpha+1) &= \Big(\Spin^2 - S_\alpha^3 S_{\alpha+1}^3 -
\Delta^{-1} (S_\alpha^1 S_{\alpha+1}^1 + S_\alpha^2 S_{\alpha+1}^2)\\ 
&\qquad \qquad + \Spin \sqrt{1 - \Delta^{-2}} (S_\alpha^3 - S_{\alpha+1}^3) \Big)\, .
\end{split}
\end{equation}
In comparison to the Hamiltonian \eq{intro hamiltonian}, we have just added the boundary
fields \eq{boundary field}, scaled by $\Delta^{-1}$ and added a constant.
One can easily check that the interaction $h^{\Spin}(\alpha,\alpha+1)$ is nonnegative.
For finite volume $\Lambda$, it is an easy but important observation that the Hamiltonian commutes with
$S^3_{\Lambda} = \sum_{\alpha \in \Lambda} S^3_\alpha$.
We use this symmetry to block diagonalize $H^{\Spin}_{\Lambda}$.
In particular, we let $\Hil(\Lambda,\Spin)$ be the spin $\Spin$ Hilbert space,
and we define $\Hil(\Lambda,\Spin,M)$ to be the eigenspace of $S^3_\Lambda$ with eigenvalue
$M \in \{-J|\Lambda|,\dots,J|\Lambda|\}$.
We call these subspaces ``sectors''.
They are invariant subspaces for $H^{\Spin}_\Lambda$.

For finite volumes $\Lambda$, the ground states of $H^{\Spin}_{\Lambda}$
may be expressed in closed form, as was pointed out in \cite{ASW}.
We will give a formula for these ground states in the next section.
For now we merely mention the fact that for each sector there is a unique ground
state $\Psi_0(\Lambda,\Spin,M)$, and it has the property that its energy is zero.
We define the ground state space
$\GS(\Lambda,\Spin,M)$ to be the one-dimensional span of $\Psi_0(\Lambda,\Spin,M)$,
then the spectral gap is given by
\begin{equation}
\label{gap defn}
\gamma(\Lambda,\Spin,M) = \inf_{\substack{\psi \in \Hil(\Lambda,\Spin,M) \\ 
\psi  \perp \GS(\Lambda,\Spin,M)}} \langle{H^{\Spin}_\Lambda}\rangle_{\psi}
\end{equation}
where $\langle{\cdots}\rangle_\psi = \ip{\psi}{\cdots \psi}/\ip{\psi}{\psi}$.

One passes to the thermodynamic limit, by considering the infinite volume
Hamiltonian as the generator of the Heisenberg dynamics on the algebra of quasilocal
observables $\Obs_0$.
The definition of a ground state is a state on $\Obs_0$ such that for any local observable
$X \in \Obs_{\Lambda}$, $|\Lambda|<\infty$, $\omega$ satisfies
\begin{equation}
\label{local stability}
\omega(X^*\delta(X)) \geq 0
\end{equation}
where $\delta(X) = \lim_{\Lambda \nearrow \Ir}[H^{\Spin}_\Lambda,X]$.
As in the case of finite volumes, there is a collection of ground states whose
GNS representation is explicit.
These ground states were discovered in \cite{GW}, and they were proven to be the complete
list in \cite{KN3}.
The infinite volume ground states are the following: 
a translation invariant up spin state determined by the equation
$\omega^\uparrow(S^3_\alpha) = +\Spin$ for all $\alpha$;
a translation invariant down spin state $\omega^\downarrow$;
an infinite number of kink states which we label $\omega^{\downarrow\uparrow}_M$;
and an infinite number of antikink states, $\omega^{\uparrow\downarrow}_M$.
The kink states have the property that, if $T$ is the translation to the left one unit 
(so $T^{-1} S_\alpha^3 T = S_{\alpha+1}^3$), then for any quasilocal observable $X$
\begin{equation}
\label{limiting behavior of kinks}
\begin{split}
\lim_{n \to \infty} \omega^{\downarrow\uparrow}_M(T^{-n} X T^n) = \omega^\uparrow(X)\, ,\\
\lim_{n \to \infty} \omega^{\downarrow\uparrow}_M(T^n X T^{-n}) = \omega^\downarrow(X)\, .\\
\end{split}
\end{equation}
The label $M$ is any integer and is determined as follows.
All the kink states are also local perturbations of one another, which we will see in the next section
when we write the explicit GNS representation.
So there is only one GNS Hilbert space for all the kink states,
and only one GNS Hilbert space for all the antikink states.

For any ground state $\omega$, the infinite volume Hamiltonian 
can be represented as the generator of
the Heisenberg dynamics for the algebra of observables on $\Hil_{\textrm{GNS}}$, 
the GNS Hilbert space of $\omega$.
This means that there is a densely defined, self adjoint operator $H_{\textrm{GNS}}$
with the property that for any $X \in \Obs_0$, 
$$
H_{\textrm{GNS}} \pi(X) \Omega_{\textrm{GNS}} = \pi(\delta(X)) \Omega_{\textrm{GNS}}\, ,
$$
where $\Omega_{\textrm{GNS}}$ is the representation of the ground state as a vector, 
$\pi$ is the representation of the quasilocal observable algebra on the observable
algebra of $\Hil_{\textrm{GNS}}$, and $\delta(X) = \lim_{\Lambda \nearrow \Ir} [H^{\Spin}_\Lambda,X]$
is the derivation defining the Heisenberg dynamics.
The bottom of the spectrum of $H_{\textrm{GNS}}$ is 0.
The spectral gap above $\omega$ is defined to be the gap (if one exists) above 0 in the spectrum
of $H_{\textrm{GNS}}$.
Note that there is one spectral gap for each of the four classes of ground states: all up, all down,
kinks, and antikinks.
 
We can now state the main result of \cite{KN1}:

\begin{theorem}\cite{KN1}
\label{Spin-1/2 spectral gap}
For the $\textrm{SU}_q(2)$ invariant spin-$1/2$ ferromagnetic XXZ chain
with the length $L \geq 2$ and $\Delta \geq 1$, the spectral gap is
$$
\gamma([1,L],1/2,M) = 1 - \Delta^{-1} \cos(\pi/L)\, ,
$$
in any sector $\Hil([1,L],1/2,M)$, $-L/2<M<L/2$.
Above any of the infinite-volume ground states (all up, all down, kink or
antikink) the spectral gap is
$$
\gamma = 1 - \Delta^{-1}\, .
$$
\QED
\end{theorem}

The previous theorem is the starting point of our own analysis.
Our main result is an analogous theorem, extending the existence of the spectral
gap to all $\Spin$ instead of just $\Spin=1/2$.

\begin{theorem}
\label{Our theorem}
For any $\Spin \in \frac{1}{2}\Nl$, 
and any $\Delta > 1$, the gap above the translation invariant ground
states is
$$
\gamma_{\textrm{up}} = \gamma_{\textrm{down}} = 2\Spin (1-\Delta^{-1})\, .
$$
The gap above the kink states satisfies the bounds 
$$
0 < \gamma_{\textrm{kink}} \leq \gamma_{\textrm{up}}\, .
$$
Specifically the spectral gap above the kink ground state is nonvanishing.
\end{theorem}

\noindent
\textit{Remark }
In the theorem, the formula for $\gamma_{\textrm{up}}$ is well-known,
The inequality $\gamma_{\textrm{kink}} \leq \gamma_{\textrm{up}}$
is also well-known and easy to deduces.
We include proofs of these facts for the convenience of the reader.
Our main result, which is new, is that $\gamma_{\textrm{kin}}$ is strictly positive.

We can say something more specific about the low excitation spectrum by considering
an extra symmetry of the Hamiltonian.
Since $H^{\Spin}_{\Lambda}$ commutes with $S^3_\Lambda$ for each finite volume $\Lambda$, we would like
to define an infinite volume analogue of $S^3_\Lambda$.
For the GNS space above the kink ground states the correct definition is the following
renormalized version
$$
\widetilde{S}^3 = \sum_{\alpha \in \Ir} (S^3_\alpha - \textrm{sign}(\alpha - 1/2) \Spin)\, ,
$$
which is a densely defined self adjoint operator on the $\Hil_{\textrm{GNS}}$.
The ground state space of $H_{\textrm{GNS}}$ is spanned by the orthogonal family of vectors
$\{\Psi_0(\Ir,\Spin,M) : M \in \Ir\}$ which are determined up to scalar multiplication
by the properties that
$$
H_{\textrm{GNS}} \Psi_0(\Ir,\Spin,M) = 0\, ,\quad
\widetilde{S}^3_\Ir \Psi_0(\Ir,\Spin,M) = M \Psi_0(\Ir,\Spin,M)\, .
$$
We define a version of the spectral gap for Hamiltonian restricted to the sectors of $\widetilde{S}^3$ 
in the following way.
Let $\gamma(\Ir,\Spin,M)$ be the largest number such that for any local observable
$X \in \Obs_\Lambda$ commuting with $S^3_\Lambda$ we have
\begin{align*}
&\ip{\Psi_0(\Ir,\Spin,M)}{\pi(X)^* H_{\textrm{GNS}}^3 \pi(X) \Psi_0(\Ir,\Spin,M)} \\
&\qquad \qquad  \geq \gamma(\Ir,\Spin,M) 
  \ip{\Psi_0(\Ir,\Spin,M)}{\pi(X)^* H_{\textrm{GNS}}^2 \pi(X) \Psi_0(\Ir,\Spin,M)}\, .
\end{align*}
An arbitrary local observable does not commute with $S^3_\Lambda$.
However, one may define $X_M$ for $-\Spin |\Lambda| \leq M \leq \Spin |\Lambda|$ so
that each $X_M$ commutes with $S^3_\Lambda$ and 
$\pi(X) \Psi_0(\Ir,\Spin,M) = \sum_{M'} \pi(X_{M'}) \Psi_0(\Ir,\Spin,M+M')$.
This is just due to the fact that the GNS representation $(\Hil_{\textrm{GNS}},\pi,\Psi_0(\Ir,\Spin,M))$
is cyclic for any choice of $M$.
From this we see that
$$
\gamma_{\textrm{kink}} = \inf_{M \in \Ir} \gamma(\Ir,\Spin,M)\, .
$$

Let $T$ be translation to the left, as before.
We have $T^{-1} \widetilde{S}^3 T = 2 \Spin + \widetilde{S}^3$, 
which implies
$$
T \Psi_0(\Ir,\Spin,M) = \Psi_0(\Ir,\Spin,M+2\Spin)\, ,
$$
since $T$ clearly commutes with the Hamiltonian.
Hence, 
$$
\gamma(\Ir,\Spin,M) = \gamma(\Ir,\Spin,M+2\Spin)\, .
$$
Another symmetry of the Hamiltonian is obtained by taking a 
left-right reflection of the lattice about the origin,
and simultaneously flipping the spin at every site.
This is a unitary transformation of $\Hil_{\textrm{GNS}}$ to itself.
Calling this symmetry $\mathcal{R}$ we have
$\mathcal{R} \widetilde{S}^3 \mathcal{R} = - \widetilde{S}^3$.
So 
$$
\gamma(\Ir,\Spin,M) = \gamma(\Ir,\Spin,-M)\, .
$$
To prove that the gap above the infinite volume kink states is nonzero,
it suffices to check that
$$
\gamma(\Ir,\Spin,M) > 0 
$$
for $M=0,1,\dots,\ceil{\Spin}$, where $\ceil{\Spin}$ is the least
integer greater than $\Spin$.
This fact does not actually simplify the proof, but the
symmetries above are an important part of our proof.

\begin{figure}[t]
\begin{center}
\epsfig{file=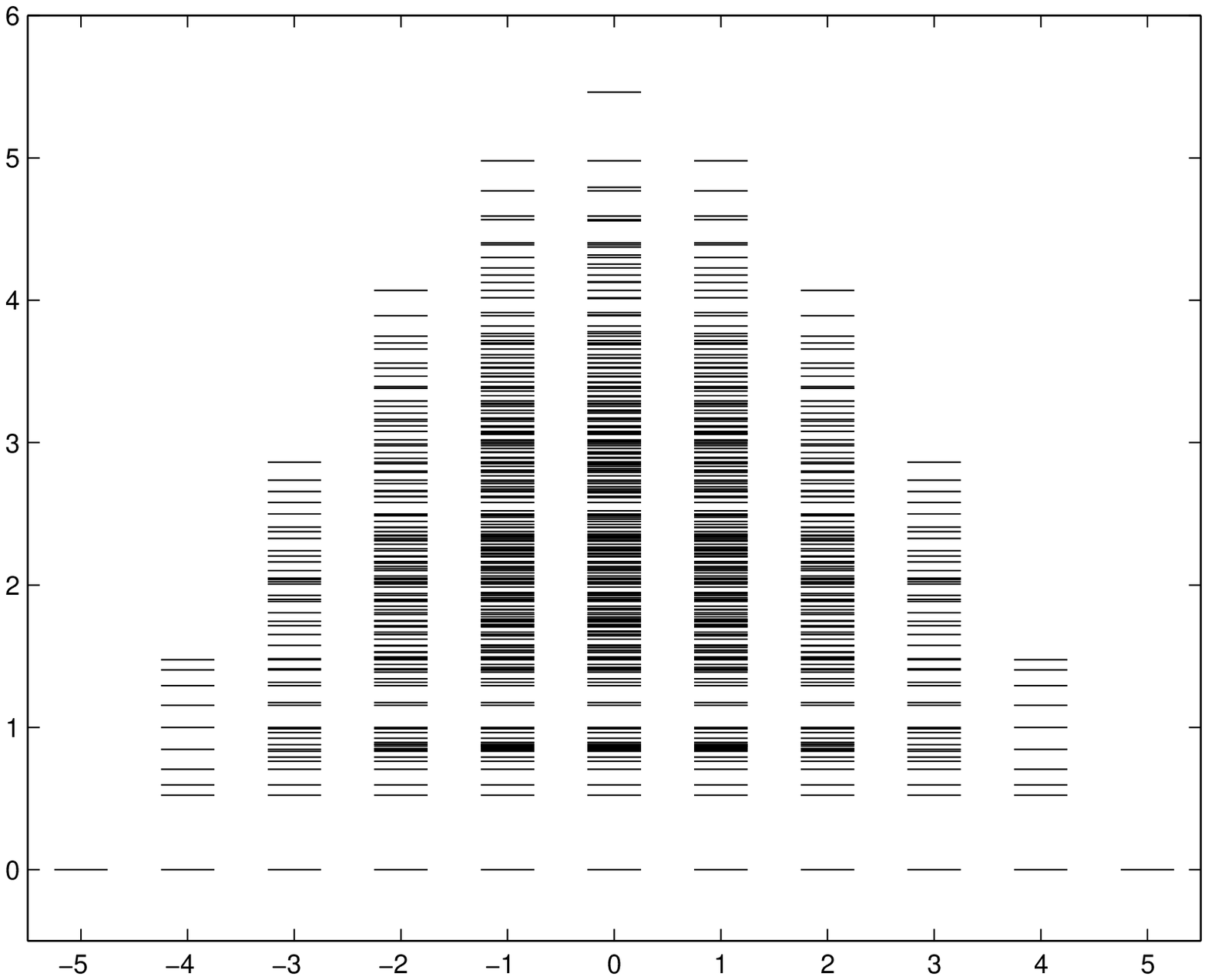, height=6cm, width=6cm}
\epsfig{file=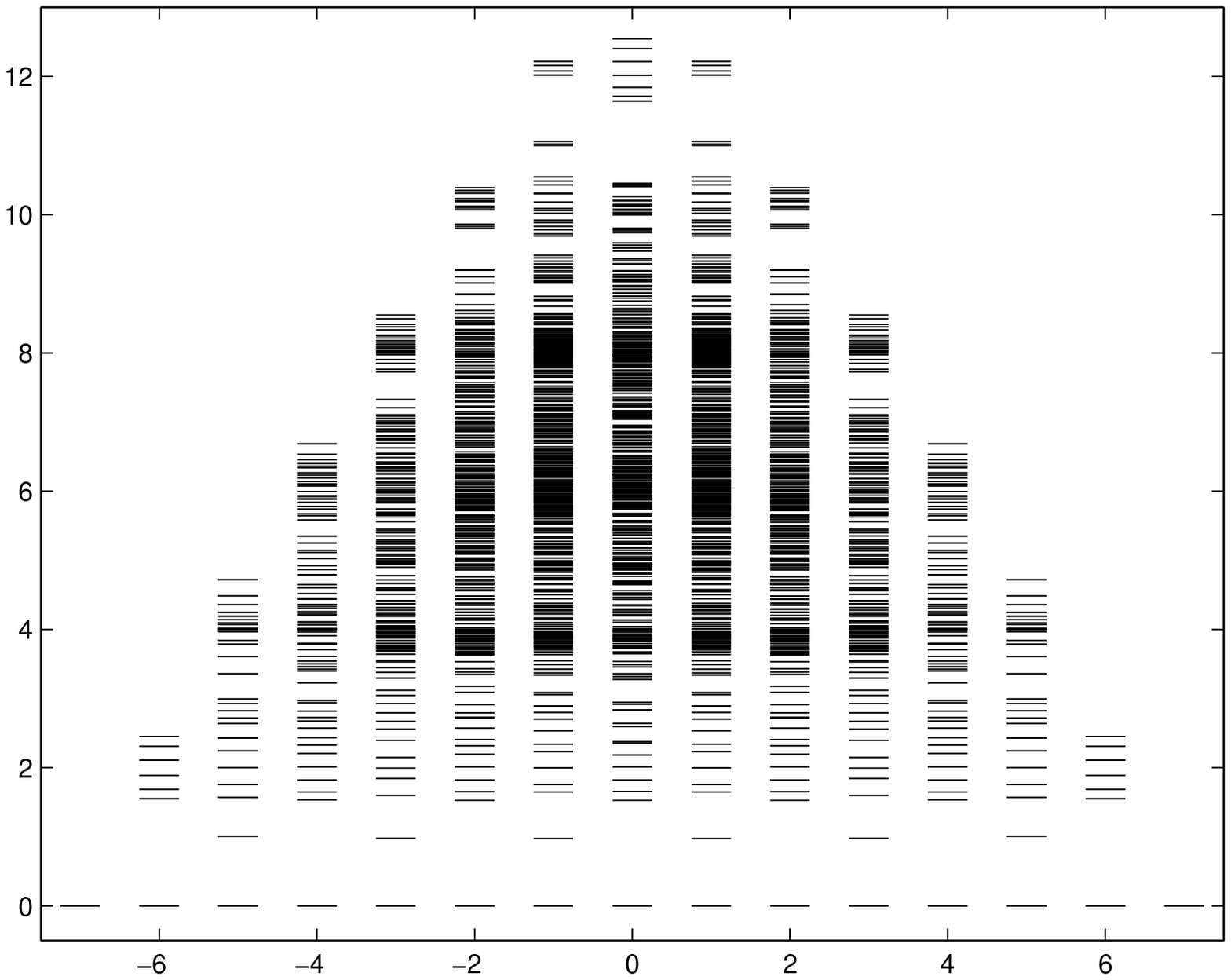, height=6cm, width=6cm}\\
\epsfig{file=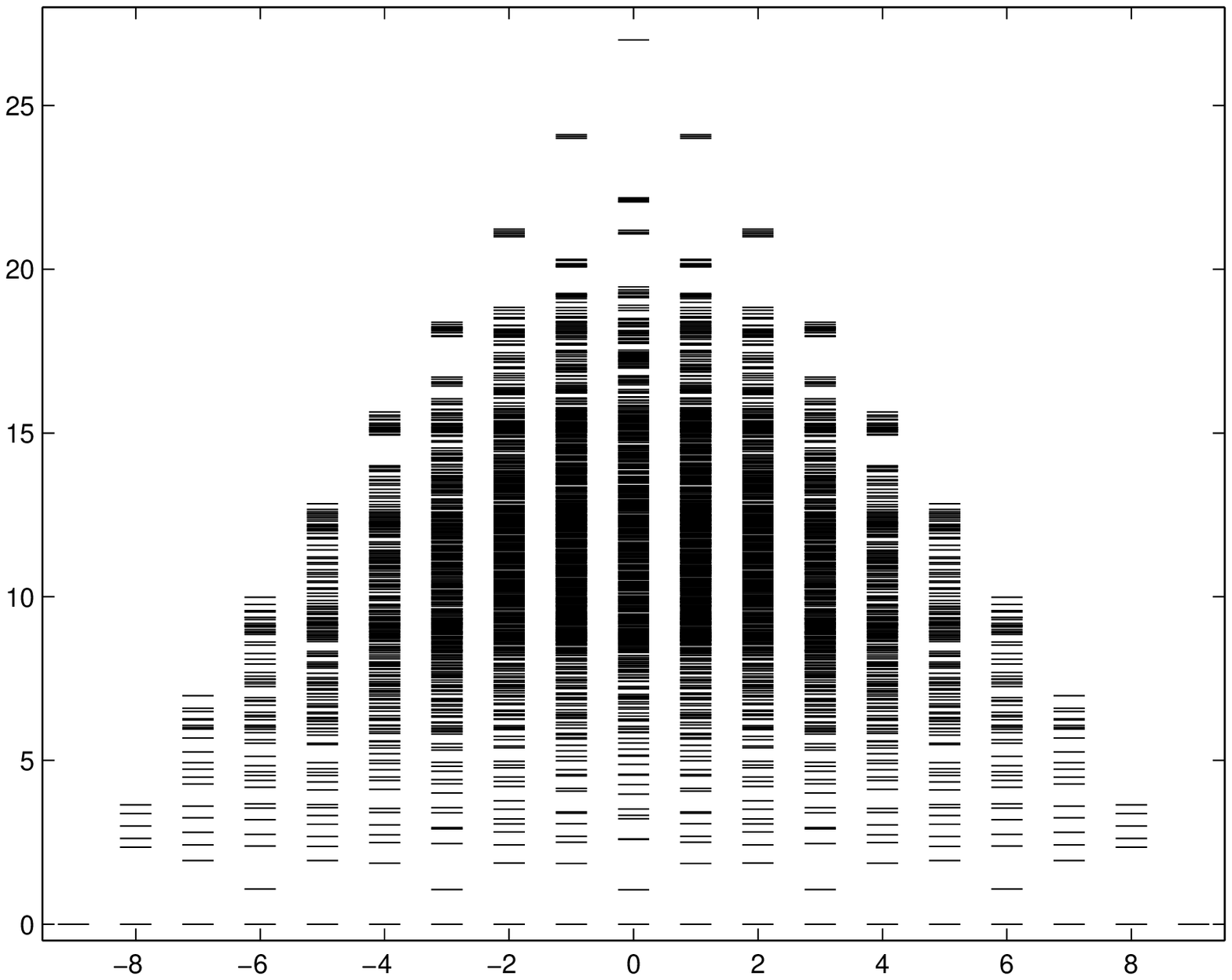, height=6cm, width=6cm}
\epsfig{file=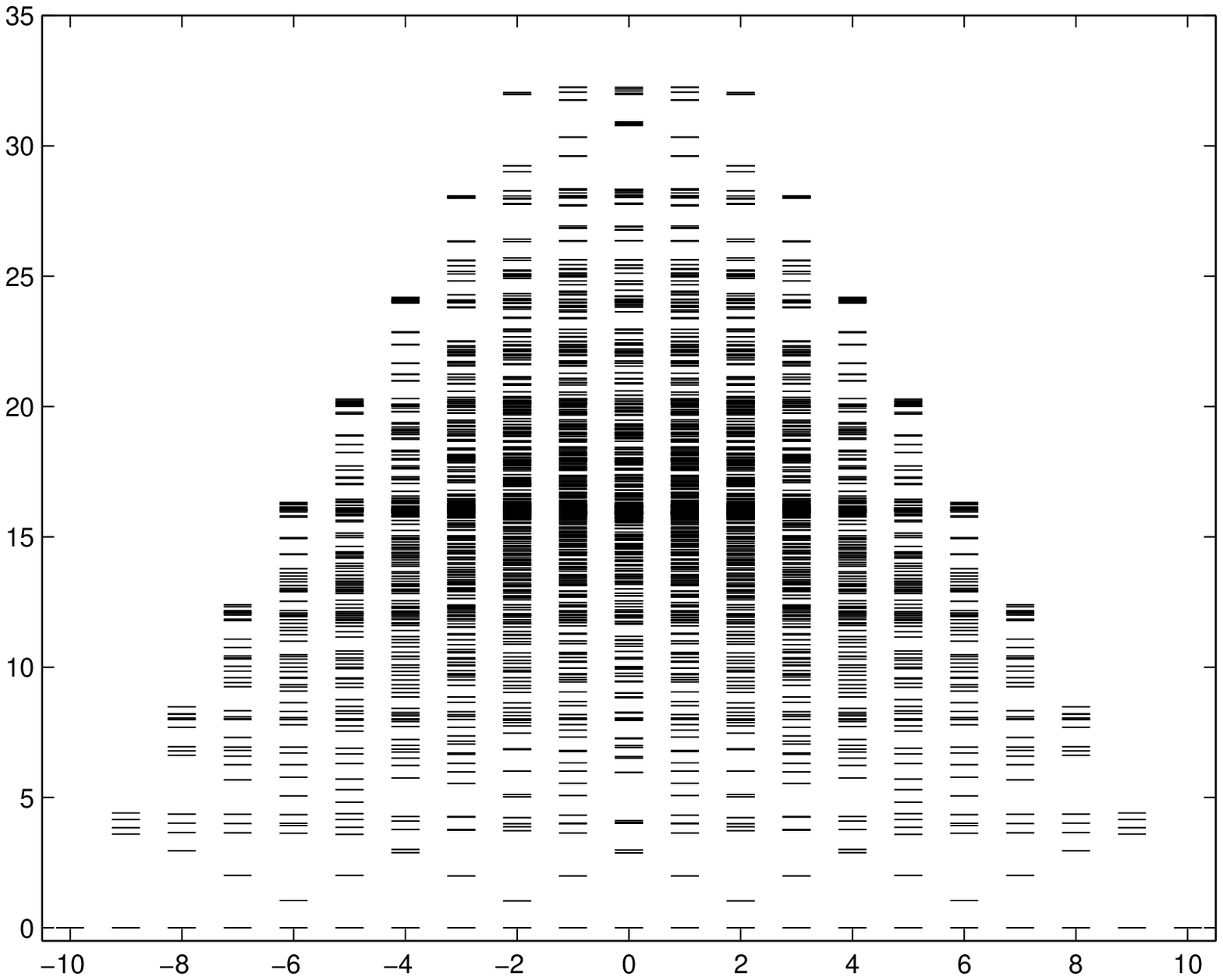, height=6cm, width=6cm}
\end{center}
\caption{
\label{gap array}
The spectrum of the XXZ spin chain for various finite spin chains.
The horizontal axis is the total magnetization $S^3_{\textrm{tot}}$ of a sector:
the lines above each number show the eigenvalues in that sector. 
The spin systems are, from left to right and top to bottom, 
$(\Spin,L,\Delta) = (1/2,10,2)$; $(1,7,4)$;
$(3/2,6,4)$; $(2,5,8)$.}
\end{figure}

In Figure \ref{gap array} we show the spectrum for some small spin chains,
as calculated by Lanczos iteration.
Even though the lengths are finite,  one sees that the gap is an even function of $M$, 
and is nearly periodic of period $2\Spin$.
As one takes $L\to \infty$, the gap for any finite region of $M$ values is periodic
and even.

Our main theorem proves existence of a spectral gap, but it is obviously just
as interesting to know what the gap is. Unfortunately, the most information we
can gain from our proof is that the spectral gap can be well approximated by
calculating the gap in a finite volume, $L$, with an error which decreases like
$q^L$. This still leaves the problem of calculating the gap in a finite volume
$L \propto 1/\eta$. We do not have any rigorous bounds for the spectral gap
valid for all $q\in (0,1)$. However we have studied the problem in three ways:
numerically, by perturbation series, and asymptotically; and we propose the
following conjectures based on our findings.

\subsection{Numerical results}
\begin{figure}
\begin{center}
\resizebox{9truecm}{9truecm}{\includegraphics{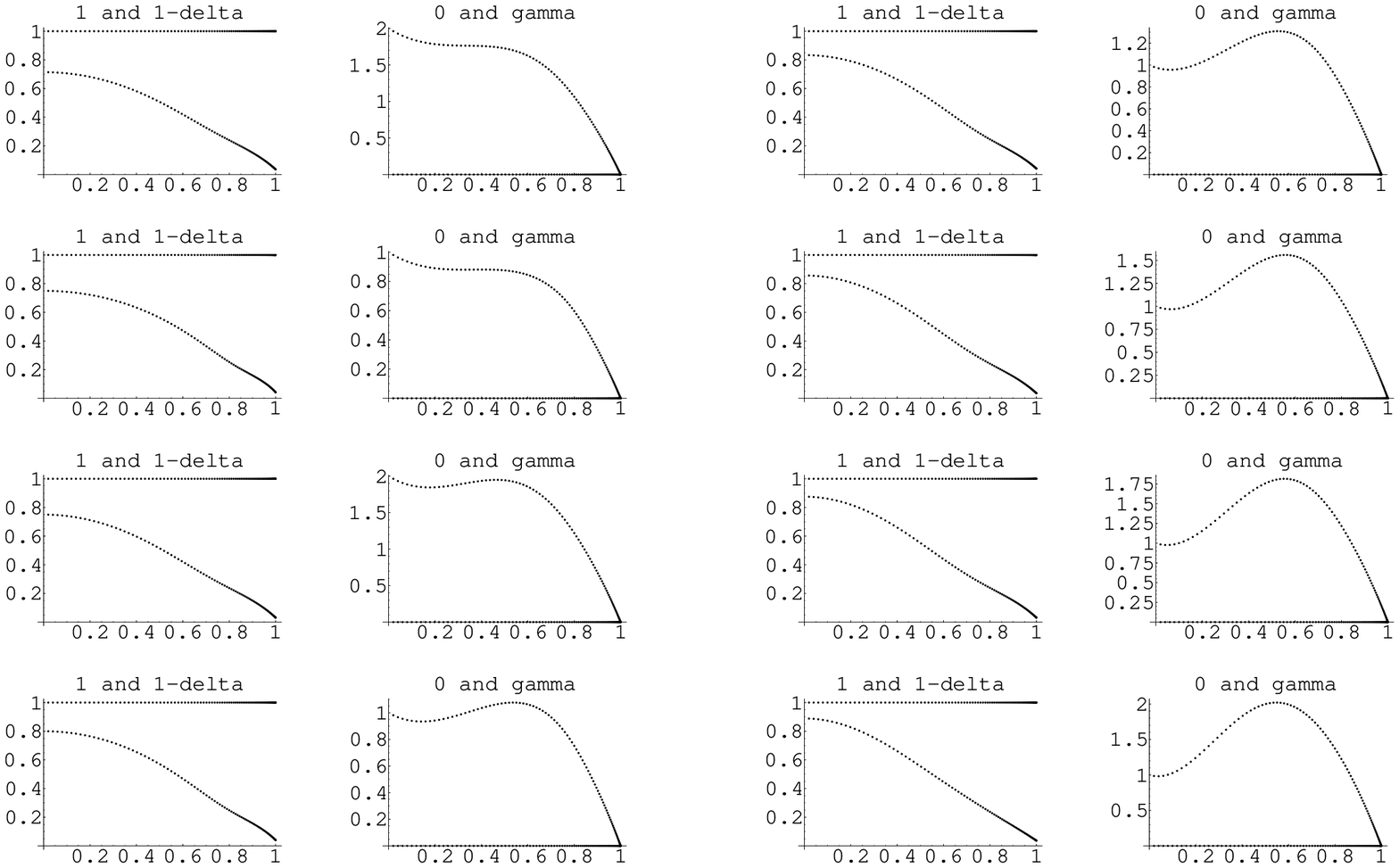}}\qquad
\end{center}
\caption{
\label{lbarray3}
Plot of $1$, $1-\delta$ (left) and $0$, $\gamma$ (right) versus
$\Delta^{-1}$, for $(\Spin,n,L)$ equal to: First column  
$(7/2, 1, 4)$, $(2, 0, 6)$, $(4, 1, 4)$, $(5/2, 0, 5)$;
Second column $(3, 0, 4)$, $(7/2, 0, 4)$, $(4, 0, 4)$, $(9/2, 0, 3)$}
\label{lowerbound3}
\end{figure}
\begin{figure}
\begin{center}
\resizebox{5truecm}{5truecm}{\includegraphics{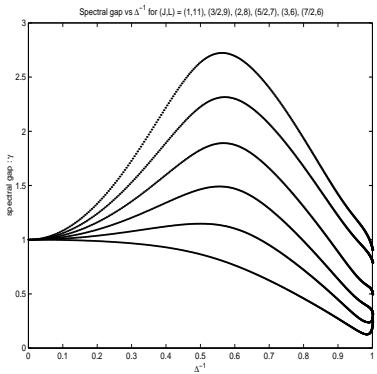}}\qquad
\end{center}
\caption{
\label{lanpics}
Some plots of the spectral gap using Lanczos iteration. For all three curves,
$n\equiv 0 \mod 2J$.}
\label{lanczospics}
\end{figure}

We performed two types of numerical methods.
The first, and more efficient method is based on the spin ladder reduction
from our proof.
In Theorem \ref{Spin ladder bound} below, we obtain
a rigorous lower bound for $\gamma([1,L],\Spin,M)$ as
$$
\gamma([1,L],\Spin,M) \geq 2\Spin (1-\Delta^{-1}) (1 - \delta([1,L],\Spin,M))
$$
where $1-\delta([1,L],\Spin,M)$ is the spectral gap for a reduced model
resembling a quantum solid-on-solid model for the spin ladder.
What is important about the bound is that the reduced model has dimension
$(L+1)^{2\Spin}$ as opposed to the original system with dimension $(2\Spin+1)^L$.
We then numerically diagonalized the reduced system to find the spectral gap for
some values of $\Spin$ and $L$.
The second numerical method was simply to numerically diagonalize the original Hamiltonian
for some small values of $L$ and $\Spin$.
We did this primarily to check the qualitative results of the lower bound.
In Figure \ref{lbarray3} we show the results of the lower bound calculation,
and in Figure \ref{lanpics} the result of the Lanczos iteration.
What emerges qualitatively is that for $\Spin>1$ there is a local maximum for the
spectral gap with $1<\Delta<\infty$.
This is other than expected based on the spin 1/2 results. 
Based on our numerical evidence we make the following conjecture.

\begin{conjecture}
\label{Numerical conjecture}
We have defined $\gamma(\Ir,\Spin,M)$ above for fixed $\Delta$.
Let us rewrite this as $\gamma(\Ir,\Spin,M,\Delta^{-1})$ to take account
of the anisotropy $0\leq \Delta^{-1} \leq 1$.
We conjecture that 
$$
\min_M \gamma(\Ir,\Spin,M,\Delta^{-1}) = \gamma(\Ir,\Spin,0,\Delta^{-1})\, ,
$$
and that for each $\Spin$ there exists a $\Delta_\Spin^{-1}$ such that 
$$
\gamma(\Ir,\Spin,0,\Delta^{-1}) \leq \gamma(\Ir,\Spin,0,\Delta_\Spin^{-1})
$$
whenever $0\leq\Delta^{-1} \leq 1$, equality holding only if $\Delta^{-1}=\Delta_\Spin^{-1}$.
For $\Spin=1/2$, the conjecture is a known fact following from
Proposition \ref{Spin-1/2 spectral gap},
and one sees $\Delta_{1/2}^{-1} = 0$. 
We conjecture that $\Delta_1^{-1}=0$ as well,
but that, for $\Spin \geq 3/2$, $0 < \Delta_\Spin^{-1} < 1$.
\end{conjecture}

\subsection{Ising Perturbation}
If we write the Hamiltonian as an expansion in $\Delta^{-1}$,
we can make a perturbation expansion off of the Ising model.
The Ising model spectrum is well known, but some interesting
facts arise.
One fact which is useful is that the Hamiltonian obtained by 
changing $\Delta^{-1}$ to $-\Delta^{-1}$ is unitarily equivalent
to the original: just rotate every other site by $\pi$ about $S^3$.
This means that $\Delta^{-1}=0$ is either a local minimum
or a local maximum of $\gamma(\Ir,\Spin,M,\Delta^{-1})$.
For $\Spin=1/2$, the first excitation above a kink ground state in the Ising limit
is infinitely degenerate.
This is why the slope of $\gamma(\Ir,1/2,M,\Delta^{-1})=1-\Delta^{-1}$ is not zero
at $\Delta^{-1}=0$.
Similarly for $\Spin=1$ and $M$ odd.
However, for all other choices of $\Spin$ and $M$, the first excitations are at most
finitely degenerate, and for $\Spin>1$ and $M=0$ the first excitations are nondegenerate.
This means that
the first derivative of $\gamma(\Ir,\Spin,M,\Delta^{-1})$ vanishes for all other
values of $\Spin$ and $M$.
For $\Spin=1$, $M=0$ the second derivative is negative, while
for $\Spin>1$ and $M=0$ the second derivative is always positive, indicating that
the Ising limit does not maximize the spectral gap, but minimizes it locally.
This argument, which will be expanded in Section 6, illuminates part of
Conjecture \ref{Numerical conjecture}.
The nondegeneracy of the first excitations suggests a second gap.
Based in part on this evidence, we make the following conjecture:
\begin{conjecture}
\label{Ising conjecture}
1)
For $1<\Delta<\infty$, $\Spin \geq 3/2$ and any $M \in \Ir$,
the lowest excited state is an isolated eigenvalue, i.e.\ there is a
nonvanishing gap to the rest of the spectrum.\\
2) For $\Spin = 1$ and $M$ any odd integer, the lowest excited state
is the bottom of a branch of continuous spectrum.
For $\Spin = 1$ and $M$ even, the lowest excited state is 
again an isolated eigenvalue.
\end{conjecture}

\subsection{Asymptotics}
From the numerics it became clear that for $M=0$ and $\Delta$ fixed, the gap 
$\gamma(\Ir,\Spin,0,\Delta^{-1})$ scales like $\Spin$.
Also, a careful analysis of the exact formula for the ground states of \cite{ASW},
which will be presented in the next section, shows that for large $\Spin$,
the wave vector has Gaussian fluctuation on the order of $J^{1/2}$.
These two facts together suggest a scaling analysis of the bottom of the
spectrum of $H_\Lambda^\Spin$ in the limit $\Spin \to \infty$.
Consistent with the Gaussian form for the ground states, our asymptotic
analysis leads to a free Boson gas model for the bottom of the spectrum,
at least to first order in $J^{-1/2}$.
We derive a boson model for the XXZ spin system analogous to \cite{HP}.
(See also \cite{Dys1,Dys2} for a better introduction to spin waves.
Unfortunately our treatment is not as well developed.)
This is not the same analysis as was done
in \cite{L4}, nor in any other coherent states approach. 
We analyze the eigenstates whose energy scales like $\Spin$,
whereas coherent states give rigorous bounds on the bulk spectrum which
scales like $\Spin^2$.
The higher energy states are much greater in number, so typically they control the
thermodynamic behavior.
However, the low energy states may be more important for dynamical properties (cf \cite{Dys1}),
since that part of the spectrum is separated by spectral gaps.
Also note, the asymptotic model is that of a free Bose gas, 
with a nontrivial dispersion relation for the energy of each oscillator.
In fact, the energy is such that excitations which give least energy are exponentially localized
about the interface.
We believe that this makes it plausible to prove the Boson gas estimate is
correct with small errors for large but fixed $\Spin<\infty$, because even as one takes
$L \to \infty$, the low excitations are ``essentially finite'' and we can more or less
prove the large $\Spin$ asymptotics for the finite system.
Of course $\Spin$ would have to depend on the number of excitations that you wanted
to estimate by the Bose gas picture; as the number of excitations goes to $\infty$
so must $\Spin$.
Our boson model has a quadratic coupling, but a Bogoliubov transformation
diagonalizes it.
The spectrum of the coupling matrix, then gives the value for the limiting curve of
the spectral gap, and other information about the low spectrum.
Based on our analysis we make the following conjecture:

\begin{conjecture}
\footnote{
\textbf{Note Added in Proof:}
Since the submission
of this paper to the arXiv, Caputo and Martinelli have obrtained further
results which prove part of Conjecture 2.5. In \cite{CapMart2} they obtain a
lower bound for the gap of the form $\Spin \times \textrm{constant}$.}
There is a function 
$\gamma_{\infty} : (1,\infty) \times \Rl \to \Rl$ with the property that
$$
\lim_{\Spin \to \infty} \Spin^{-1} \gamma(\Ir,\Spin,\mu \Spin,\Delta) 
  = \gamma_{\infty}(\mu,\Delta)\, .
$$
This function satisfies 
$$
\gamma_\infty(\mu+2,\Delta) = \gamma_\infty(\mu,\Delta)\qquad
\textrm{and}\qquad 
\gamma_\infty(-\mu,\Delta) = \gamma_\infty(\mu,\Delta)\, .
$$
Moreover $\gamma_\infty(\mu,\Delta)$ is equal to
the spectral gap of a bi-infinite Jacobi operator $A$ (really $A(\Delta,\mu)$) defined 
on $l^2(\Ir)$
\begin{equation}
\label{sgJacobi}
\begin{split}
A e_n &= [2 e_n - \sech(\eta) e_{n-1} - \sech(\eta) e_{n+1}] \\
  &- \frac{4 \sinh^2(\eta)}{\cosh(2 \eta(n-r)) + \cosh(2 \eta)}\, e_n\, .
\end{split}
\end{equation}
Above, $r = r(\mu,\Delta)$ is a phase defined implicitly by the equation
$$
\mu = \lim_{n\to\infty} \sum_{-n+1}^n \tanh(\eta(n-r)) \, .
$$
\end{conjecture}

\medskip
\textit{Remark 1.}
The implicit formula for $\gamma_\infty(\mu,\Delta)$ has a specific consequence
that
$$
\lim_{\Spin \to \infty} \Delta_{\Spin}^{-1} = 0.49585399 \pm 10^{-8}\, ,
$$
which is obtained by numerical diagonalization of large Jacobi matrices.

\medskip
\textit{Remark 2.} Equation \eq{sgJacobi} has a simple physical interpretation.
The bracketed term on the RHS is the usual matrix for a one-magnon spin wave,
which by itself would give the energy $2(1-\sech \eta)$ for the spectral gap,
just as in the translation invariant ground states.
The second term on the RHS of \eq{sgJacobi} represents the attractive potential
due to the domain wall with center $r$. 

\begin{figure}[t]
\begin{center}
\epsfig{file=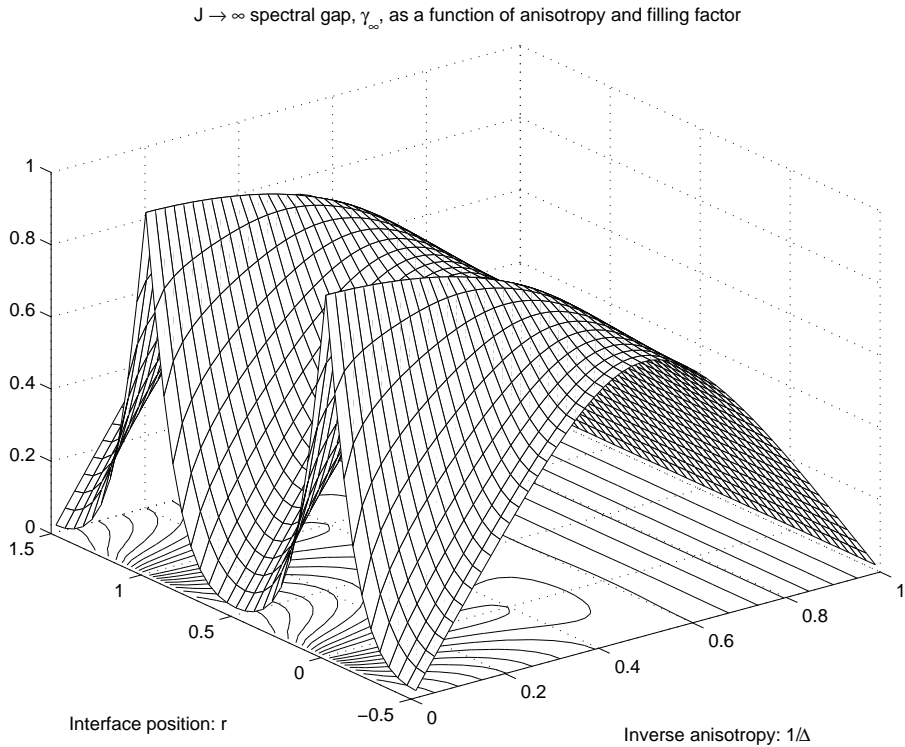, height=10cm}
\end{center}
\caption{
\label{sgpic}
Surface and contour plot for the function $\gamma_\infty$ versus $r$ and $\Delta^{-1}$}
\end{figure}

\medskip
In Figure \ref{sgpic} we show the function $\gamma_\infty(r,\Delta^{-1})$
as obtained by numerically diagonalizing the Jacobi operator for 50 sites.
Note that the number of sites reflects the extreme simplicity of the Bose
model over the true XXZ model; we could not numerically diagonalize
a spin chain of 50 sites even for spin $1/2$.

\Section{Ground states of the XXZ model}
In this section we give formulas for the ground states of the 
finite volume XXZ model as in \cite{ASW}.
We show how the ground states of the spin $\Spin$ chain can be
derived by looking at the ground states of a spin $1/2$ ladder with
$2\Spin$ legs, which is essential for our proof.
We also recall the explicit GNS representation of the infinite volume ground states
in a Guichardet Hilbert space (also called incomplete tensor product) \cite{GW,KN3}.
This concrete representation is convenient, especially for proving that certain sequences
of finite volume ground states have unique limits.

We begin by rewriting \eq{Hamiltonian definition} for spin $1/2$.
It is easy to check that for the two site interaction $h^{1/2}(1,2)$,
the three vectors 
$$\ket{+\frac{1}{2},+\frac{1}{2}}\, ,\quad
\ket{-\frac{1}{2},-\frac{1}{2}}\, ,\quad 
\ket{-\frac{1}{2},+\frac{1}{2}} + q \ket{+\frac{1}{2},-\frac{1}{2}}
$$
are ground states, while $\ket{+1/2,-1/2} - q \ket{-1/2,+1/2}$ is a state with energy one.
In other words, we may view $h^{1/2}(1,2)$ as $\unity - U(\tau(1,2))$, where 
$\tau(1,2) \in \mathfrak{S}_2$ is the transposition, and $U$ is the 
(non-unitary) action of $\mathfrak{S}_2$ defined by
$$
U(\tau(1,2))\phi(m_1,m_2)=\phi(m_2,m_1),
$$
where the $\phi(m_1,m_2)$ are the (non-normalized) basis vectors 
\begin{alignat*}{3}
\phi(\frac{1}{2},\frac{1}{2}) &= \ket{\frac{1}{2},\frac{1}{2}}\, , 
&\quad \phi(\frac{1}{2},-\frac{1}{2}) &= q^{-1/2} \ket{\frac{1}{2},-\frac{1}{2}}\, ,\\
\phi(-\frac{1}{2},-\frac{1}{2}) &= \ket{-\frac{1}{2},-\frac{1}{2}}\, , 
&\quad \phi(-\frac{1}{2},\frac{1}{2}) &= q^{1/2} \ket{-\frac{1}{2},\frac{1}{2}}\, .
\end{alignat*}
In other words, the ground states of the two site Hamiltonian are the symmetric
tensors with respect to the nonunitary action $U$.
We can generalize this result to linear chains of any length,
and to many other domains as well.
Specifically, what we need to properly define the XXZ Hamiltonian with boundary fields
is a collection of sites $\Lambda$ and a collection of oriented
bonds among those sites $\mathcal{B}$, in other words a digraph.
Then we define
$$
H^{\Spin}_{\Lambda,\mathcal{B}} = \sum_{(\alpha,\beta) \in \mathcal{B}} h^{\Spin}(\alpha,\beta)\, .
$$
(We write $H^{\Spin}_{\Lambda}$ when $\mathcal{B}$ is obvious.)
We define a height function to be any function $l : \Lambda \to \Ir$ such that
$l(\beta) - l(\alpha) = 1$ for all $(\alpha,\beta) \in \mathcal{B}$.
The condition to have such a height function is that for any closed loop, where a loop is
defined as a sequence $\alpha_1, \alpha_2,\dots, \alpha_n=\alpha_1 \in \Lambda$
such that for each $i$ either $(\alpha_i,\alpha_{i+1}) \in \mathcal{B}$ or
$(\alpha_{i+1},\alpha_i) \in \mathcal{B}$, there are equal numbers of bonds with positive
orientation $(\alpha_i,\alpha_{i+1}) \in \mathcal{B}$ as with negative orientation
$(\alpha_{i+1},\alpha_i) \in \mathcal{B}$.
The following lemma is an interpretation of a result in \cite{ASW}.
\begin{lemma}
If $(\Lambda,\mathcal{B})$ is a connected digraph such that a height function
$l$ exists, then there is a unique ground state of $H^{1/2}_{\Lambda,\mathcal{B}}$
in each sector $\Hil(\Lambda,1/2,M)$ for $M \in [-|\Lambda|/2,|\Lambda|/2]$.
\end{lemma}

\begin{proof}
The proof is like the analogous statement (without the requirement of a height function)
for the isotropic model.
Suppose that $l$ exists.
Define a (non-normalized) basis of vectors
$$
\phi(\{m_\alpha\}) = q^{-\sum_{\alpha \in \Lambda} m_{\alpha} l(\alpha)} \ket{\{m_\alpha\}}\, .
$$
Then define an action of $\mathfrak{S}_{\Lambda}$ on $\Hil(\Lambda,1/2,M)$
by
$$
U(\pi) \phi(\{m_{\alpha}\}) = \phi(\{m_{\pi^{-1}(\alpha)}\})\, .
$$
As we have already seen, for any $(\alpha,\beta) \in \mathcal{B}$, 
$h^{1/2}(\alpha,\beta) = \unity - U(\tau(\alpha,\beta))$, where $\tau(\alpha,\beta)$
is the transposition.
Thus, the unique ground state vectors are those vectors which are symmetric under the action
of $U(\mathfrak{S}_\Lambda)$, 
$$
\Psi_0(\Lambda,1/2,M) = \sum_{\substack{\{m_\alpha\} \in [-1/2,1/2]^{\Lambda} \\
  \sum_\alpha m_\alpha = M}} \phi(\{m_\alpha\})\, ,
$$
one for each sector.
\end{proof}
We note that one can trivially prove the converse of this lemma, that if there exist
ground state vectors in any sector other than $M = \pm |\Lambda|/2$, then there is a height
function (as long as $0<q<1$).
We now mention a second lemma (which is also implicit in \cite{ASW})
which gives the construction of the spin $\Spin$ ground states by using spin ladders.
\begin{lemma}
\label{Spin ladder intro}
Suppose that $(\Lambda,\mathcal{B})$ satisfies the hypotheses of the last lemma.
Then for any spin $\Spin \in \frac{1}{2} \Nl$,
there is a unique ground state of $H_{\Lambda,\mathcal{B}}^{\Spin}$ in each sector
$\Hil(\Lambda,\Spin,M)$.
Moreover, this ground state is associated to the ground state of the spin $1/2$ spin ladder
\begin{align*}
\widetilde{\Lambda} &= \{(\alpha,j) : \alpha \in \Lambda, j \in [1,2\Spin]\}\, , \\
\widetilde{\mathcal{B}} &= \{((\alpha,j),(\beta,k)) : (\alpha,\beta) \in \mathcal{B}, (j,k) \in [1,2\Spin]^2\}\, .
\end{align*}
Define $Q_\alpha : \Hil(\{\alpha\}\times[1,2\Spin],1/2) \to \Hil(\{\alpha\},\Spin)$ to be the 
projection onto the unique highest spin representation
in the decomposition of $\Hil(\{\alpha\}\times[1,2\Spin],1/2)$ into irreducibles, so that
$Q_\alpha^* Q_\alpha$ is the projection onto symmetric tensors.
Then $Q_\Lambda = \prod_{\alpha \in \Lambda} Q_\alpha$ gives an isomorphism of ground states
\begin{align*}
\Psi_0(\Lambda,\Spin,M) = Q_\Lambda \Psi_0(\widetilde{\Lambda},1/2,M)\, ,\quad
\Psi_0(\widetilde{\Lambda},1/2,M) = Q_\Lambda^* \Psi_0(\Lambda,\Spin,M)\, .
\end{align*}
\end{lemma}
\begin{proof}
Most of the proof is self evident, the point being just to introduce the notation necessary for later work.
We note that we can define a height function $\tilde{l}$ by $\tilde{l}(\alpha,j) = l(\alpha)$.
Then the ground states of $H^{1/2}_{\widetilde{\Lambda},\widetilde{\mathcal{B}}}$
are in the range of $Q_\Lambda^* Q_\Lambda$
because for any permutation $\pi$ which preserves the rungs of the ladder, i.e.\
$\pi(\{\alpha\}\times[1,2\Spin]) = \{\alpha\} \times [1,2\Spin]$ for all $\alpha$, 
$U(\pi)$ coincides with the normal action.
So
\begin{equation}
\label{turnaround}
Q_\Lambda^* Q_\Lambda \Psi_0(\widetilde{\Lambda},1/2,M) = \Psi_0(\widetilde{\Lambda},1/2,M)\, .
\end{equation}
On the other hand $Q_\Lambda H^{1/2}_{\widetilde{\Lambda}} Q_{\Lambda}^*$
equals $H^{\Spin}_{\Lambda}$.
So $Q_\Lambda \Psi_0(\widetilde{\Lambda},1/2,M)$ is a ground state of $H^{\Spin}_\Lambda$,
in fact all the ground states are obtained like this by an easy Perron-Frobenius argument.
Rewriting \eq{turnaround} with $\Psi_0(\Lambda,\Spin,M)$ in place of 
$Q_\Lambda \Psi_0(\widetilde{\Lambda},1/2,M)$ finishes the proof.
\end{proof}
In the particular case of $\Lambda \subset \Ir$, one can take $l(\alpha) = \alpha$.
Let us for future notational ease define $\mathcal{M}(\Lambda,\Spin,M)$ to be the set of all
$\{m_\alpha\} \in [-\Spin,\Spin]^{\Lambda}$ with the property that $\sum_\alpha m_\alpha = M$.
Then the formula one derives for the spin $\Spin$ ground states is
$$
\Psi_0(\Lambda,\Spin,M) = \sum_{\{m_\alpha\} \in \mathcal{M}(\Lambda,\Spin,M)}\, 
  \prod_{\alpha \in \Lambda} \binom{2\Spin}{\Spin+m_\alpha}^{1/2} q^{-\alpha m_\alpha}\ \ket{\{m_\alpha\}}\, .
$$

We now turn our attention to the infinite volume ground states.
We will merely define the ground states, the proof of completeness was done in \cite{KN3}.
Given a countably infinite set of sites $\Lambda_\infty$, and a finite dimensional Hilbert space
$\Hil_\alpha$ and unit vector $\Omega_\alpha$ at each site, one can define the Guichardet
Hilbert space
\begin{align*}
\bigotimes_{\alpha \in \Lambda_\infty} (\Hil_\alpha,\Omega_\alpha)
  &= \textrm{cl}\Bigg( \bigoplus_{n=1}^\infty \Big[ (\bigotimes_{j=1}^n \Hil_{\alpha_j} 
  \otimes \bigotimes_{j=n+1}^\infty \Cx\Omega_{\alpha_j}) \\
  &\qquad \qquad \cap
  (\bigotimes_{j=1}^{n-1} \Hil_{\alpha_j} 
  \otimes \bigotimes_{j=n}^\infty \Cx\Omega_{\alpha_j})^\perp \Big] \Bigg)\, ,
\end{align*}
where $\alpha_1,\alpha_2,\dots$ is any enumeration of $\Lambda_\infty$,
and $\textrm{cl}$ means the usual $L^2$ closure.
If $\Lambda$ is any finite subset of $\Lambda_\infty$ then we can define
the finite dimensional Hilbert space
$\Hil_\Lambda = \bigotimes_{\alpha \in \Lambda} \Hil_\alpha$, as usual,
and an obvious inclusion
$$
i_{\Lambda,\Lambda_\infty} : \Hil_{\Lambda} \to \bigotimes_{\alpha \in \Lambda_\infty} (\Hil_\alpha,\Omega_\alpha)\, .
$$
This is the proper framework to discuss the one dimensional infinite-volume ground states of the
XXZ model, because of the following
\begin{definition}
For any $\Spin \in \frac{1}{2} \Nl$ we make the following definitions. 
Consider $\Lambda_\infty = \Nl$. 
For each $\alpha \in \Lambda_\infty$
let $\Hil_\alpha = \Cx^{2\Spin+1}$ and $\Omega_\alpha = \ket{\Spin}$.
Denote the Guichardet Hilbert space so obtained by
$$
\Hil(\Nl,\Spin,\textrm{up}) = \bigotimes_{\alpha \in \Nl} (\Hil_\alpha,\Omega_\alpha)\, .
$$
Now let $\Lambda_\infty = \Ir$ instead. For each $\alpha\in \Lambda_\infty$ let 
$\Hil_\alpha = \Cx^{2\Spin+1}$ as before, but let $\Omega_\alpha = \ket{\Spin}$ for 
$\alpha \geq 1$ and $\Omega_\alpha = \ket{-\Spin}$ for $\alpha \leq 0$.
Define
$$
\Hil(\Ir,\Spin,\textrm{kink}) = \bigotimes_{\alpha \in \Ir} (\Hil_\alpha,\Omega_\alpha)\, .
$$
\end{definition}
Then
\begin{lemma}
\label{kink state convergence}
(a) If $L_k$ is a sequence of integers with $L_k \to \infty$, and $N \in \Nl$, 
then the normalized sequence 
$$
\frac{i_{[1,L_k],\Nl} \Psi_0([1,L_k],\Spin,\Spin L_k - N)}{\| \Psi_0([1,L_k],\Spin,\Spin L_k - N)\|}
$$ 
converges in norm.\\
(b) If $M_k = L_k \Spin - 2 \Spin r_k + N$ where $N \in [0,\Spin-1]$, $r_k \in \Nl$ and
both $r_k$ and $L_k - r_k$ tend to $\infty$, then the normalized sequence
$$
\frac{T^{-r_k} i_{[1,L_k],\Ir} \Psi_0([1,L_k],\Spin,M_k)}
{\|\Psi_0([1,L_k],\Spin,M_k)\|}
$$
converges in norm, where $T$ is the translation one unit to the left.
\end{lemma}
\begin{proof}
We prove this for spin $1/2$.
Then the analogue follows for ground states of $H^{1/2}_{\widetilde{\Lambda}}$,
and by the last lemma this proves it for arbitrary spin.

For (a), note that defining
$$
\Psi_0'([1,L],1/2,N) = \sum_{1\leq \alpha_1 < \dots < \alpha_N\leq L} 
  q^{\alpha_1 + \dots + \alpha_N} S_{\alpha_1}^- \cdots S_{\alpha_N}^- \Omega_{\Nl}\, ,
$$
where $\Omega_{\Nl}$ is the all up spin vector, we have
$$
\frac{i_{[1,L],\Ir}\Psi_0([1,L],1/2,\frac{1}{2}L -N)}{\|\Psi_0([1,L],1/2,\frac{1}{2}L -N)\|}
  = \frac{\Psi_0'([1,L],1/2,N)}{\|\Psi_0'([1,L],1/2,N)\|}\, .
$$
Then (a) is proved if we prove that the sequence
$\Psi_0'([1,L],1/2,N)$ converges.
But this follows by the Monotone Convergence Theorem, thinking of the 
coefficient of
$S_{\alpha_1} \cdots S_{\alpha_N} \Omega_{\Nl}$ 
as a function $f_L(\alpha_1,\dots,\alpha_N)$.
We still need to check that the limit is finite, i.e.\ that
$$
\sum_{1\leq \alpha_1 < \dots < \alpha_N} q^{2(\alpha_1 + \dots + \alpha_N)} < \infty\, .
$$
We can evaluate the series explicitly; it is $q^{N(N+1)}/\prod_{j=1}^N (1 - q^{2j})$.

For (b), we do a similar thing. 
We define $\Omega_\Ir$ to be the vector $\bigotimes_{\alpha \in \Ir} \Omega_\alpha$.
Since $\Spin=1/2$, now $M_k = \frac{1}{2} L_k - r_k$ where $r_k, L_k - r_k \to \infty$.
Then 
$$
\frac{T^{-r_k} i_{[1,L_k],\Ir} \Psi_0([1,L_k],1/2,M_k)}
{\|\Psi_0([1,L_k],1/2,M_k)\|}
  = \frac{\Psi_0'([1-r_k,L_k-r_k],1/2,0)}{\|\Psi_0'([1-r_k,L_k-r_k],1/2,0)\|}
$$
where 
\begin{align*}
\Psi_0'([-a,b],1/2,0) &= \sum_{n=0}^\infty 
  \sum_{-a\leq \alpha_1<\dots<\alpha_n\leq 0<\beta_1 < \dots < \beta_{n}\leq b} \prod_{k=1}^n
  q^{\beta_k - \alpha_k} S_{\alpha_k}^+ S_{\beta_k}^-\ \Omega_\Ir\, .
\end{align*}
The lemma will follow by the Dominated Convergence Theorem if we prove that 
$\Psi_0'(\Ir,1/2,0)$ is summable.
This is equivalent to 
$$
Z = \sum_{n=0}^\infty 
  \sum_{\alpha_1<\dots<\alpha_n\leq 0<\beta_1 < \dots < \beta_{n}}
  q^{2[\beta_1 + \dots + \beta_{n} - (\alpha_1 + \dots + \alpha_n)]} < \infty\, .
$$
We can also evaluate this explicitly
\begin{align*}
Z &= \sum_{n=0}^\infty \frac{q^{n(n+1)}}{\prod_{j=1}^n (1-q^{2j})} \cdot \frac{q^{n(n-1)}}{\prod_{j=1}^n (1-q^{2j})} \\
  &= \sum_{n=0}^\infty \frac{q^{2n^2}}{\prod_{j=1}^n (1-q^{2j})^2} \\
  &= \frac{1}{\prod_{j=1}^\infty (1 - q^{2j})}
\end{align*}
by Heine's theorem (c.f.\ \cite{GR}).
In particular it is finite.
\end{proof}
Let 
$$
\Psi_0(\Ir,\Spin,M) = \lim_{L \to \infty} \frac{\Psi_0([-L+1,L],\Spin,M)}
  {\|\Psi_0([-L+1,L],\Spin,M)\|}\, .
$$
The limit exists by the lemma.
We claim that this vector gives an infinite volume ground state.
The representation of quasi-local observables on $\Hil(\Ir,\Spin,\textrm{kink})$
is clear: the local observables $\Obs_\Lambda$ are operators of $\Hil_\Lambda$ 
which includes
by $i_{\Lambda,\Ir}$ into $\Hil(\Ir,\Spin,\textrm{kink})$.
Take the weak-$*$ completion and we are done.
The state is then $\omega^{\textrm{kink}}_{\Spin,M}(X) = \ip{\Psi_0(\Ir,\Spin,M)}{X \Psi_0(\Ir,\Spin,M)}$.
All one needs to check is that for any local observable $X$, 
$$
\lim_{\Lambda \to \infty}
\ip{\Psi_0(\Ir,\Spin,M)}{X^* [H_{\Lambda},X] \Psi_0(\Ir,\Spin,M)} \geq 0
$$
Suppose $X \in \Obs_\Lambda$.
Define $\Lambda_1$ to be the union of $\Lambda$ with all nearest neighbors.
Then for $\Lambda' \supset \Lambda_1$, $[H_{\Lambda'},X] = [H_{\Lambda_1},X]$.
Using the frustration free property of $\Psi_0([-L,L],\Spin,M)$, we see that as soon as $\Lambda_2 \subset [-L,L]$,
we have
$$
\ip{\Psi_0([-L+1,L],\Spin,M)}{X^* [H_{\Lambda_2},X] \Psi_0([-L+1,L],\Spin,M)} \geq 0\, .
$$
Taking the limit, we see that $\omega^{\textrm{kink}}_{\Spin,M}$ is an infinite volume ground state.
We define the $\gamma(\Spin,M)$ to be the gap above zero
in the spectrum of the Hamiltonian acting on the GNS space of the ground state 
$\omega^{\textrm{kink}}_{\Spin,M}$.

There are three other classes of ground states.
The antikink ground states are the states
$\omega^{\textrm{anti}}_{\Spin,M} = \omega^{\textrm{kink}}_{\Spin,M} \circ \mathcal{F}$
where $\mathcal{F}$ is uniquely determined by the formula $\mathcal{F}(S_k^+) = S_k^-$ for all $k \in \Ir$.
There are also the all up spin states and all down spin states, which are well known
and characterized by $\omega^{\textrm{up,down}}_\Spin(S_\alpha^3) = \pm \Spin$ for all $\alpha$.
We mention that the GNS space for them is also a Guichardet Hilbert space where $\Omega^{\textrm{up,down}}_\alpha 
= \ket{\pm \Spin}$ for all $\alpha$, and $\Omega^{\textrm{up,down}}_\Ir$ is the vector representing
the ground state.
That these are all the ground states is proved by Koma and Nachtergaele \cite{KN3}.


\section{Spin ladder reduction}
We now elaborate on the spin ladder construction introduced in Lemma \ref{Spin ladder intro}.
Let us define $\widetilde{H}_{\widetilde{\Lambda}} = H_{\widetilde{\Lambda},\widetilde{\mathcal{B}}}$
introduced in the lemma.
Then, defining $\GS(\Lambda,\Spin,M)$ to be the one dimensional ground state space of $H^\Spin_\Lambda$ 
in the sector $\Hil(\Lambda,\Spin,M)$, and $\GS(\widetilde{\Lambda},M)$
to be the one dimensional ground state space of $\Hil(\widetilde{\Lambda},M)$,
the lemma tells us that 
$$
Q_\Lambda \GS(\widetilde{\Lambda},M) = \GS(\Lambda,M)\, ,\quad
Q^*_\Lambda \GS(\Lambda,M) = \GS(\widetilde{\Lambda},M) = \GS(\Lambda,M)\, .
$$
The spectral gap in the sector $\Hil(\Lambda,\Spin,M)$ is defined
$$
\gamma(\Lambda,\Spin,M) = \inf_{\substack{\psi \in \Hil(\Lambda,\Spin,M) \\
  \psi \perp \GS(\Lambda,\Spin,M)}} \frac{\ip{\psi}{H^{\Spin}_\Lambda \psi}}{\ip{\psi}{\psi}}\, .
$$
In view of the lemma, 
defining $P_\Lambda = Q_\Lambda^* Q_\Lambda$, we can rewrite the spectral gap
$$
\gamma(\Lambda,\Spin,M) = \inf_{\substack{\psi \in \Hil(\widetilde{\Lambda},M) \setminus \ker P_\Lambda \\
  \psi \perp \GS(\widetilde{\Lambda},M)}} 
  \frac{\ip{P_\Lambda \psi}{\widetilde{H}_{\widetilde{\Lambda}} P_\Lambda \psi}}{\ip{P_\Lambda \psi}{P_\Lambda \psi}}\, .
$$

We now introduce a second Hamiltonian on $\Hil(\widetilde{\Lambda},M)$ which is
$$
\widetilde{\widetilde{H}}_{\widetilde{\Lambda}} = H_{\widetilde{\Lambda},\widetilde{\widetilde{\mathcal{B}}}}\, ,
$$
where 
$$
\widetilde{\widetilde{\mathcal{B}}} 
  = \{((\alpha,j),(\beta,j)) : (\alpha,\beta) \in \mathcal{B}\, ,\ j \in [1,2\Spin]\}\, .
$$
This is clearly equivalent to $2\Spin$ disjoint copies of $H^{1/2}_\Lambda$.
So, if $\Lambda=[1,L]$, Theorem \ref{Spin-1/2 spectral gap}
guarantees that the spectral gap of $\widetilde{\widetilde{H}}_{\widetilde{\Lambda}}$ is
equal to $1 - \Delta^{-1} \cos(\pi/L) > 1 - \Delta^{-1}$.
Let us introduce some notation: Let $\Hil_0(\widetilde{\Lambda},M)$
be the ground state space of $\widetilde{\widetilde{H}}_{\widetilde{\Lambda}}$
in the sector $\Hil(\widetilde{\Lambda},M)$, and let $\Hil_{\textrm{exc}}(\widetilde{\Lambda},M)$
be its orthogonal complement in the sector $\Hil(\widetilde{\Lambda},M)$.
Then
\begin{equation}
\label{Htildetilde exc}
\inf_{\psi \in \Hil_{\textrm{exc}}(\widetilde{\Lambda},M)} 
\frac{\ip{\psi}{\widetilde{\widetilde{H}}_{\widetilde{\Lambda}} \psi}}{\ip{\psi}{\psi}} \geq 1 - \Delta^{-1}\, .
\end{equation}
Defining $P_\alpha = Q_\alpha^* Q_\alpha$ to be symmetrization in the rung $\{\alpha\} \times [1,2\Spin]$,
we observe that 
$$
P_\alpha \boldsymbol{S}_{(\alpha,j)} P_\alpha 
  = \frac{1}{2\Spin} \sum_{k=1}^{2\Spin} P_\alpha \boldsymbol{S}_{(\alpha,k)} P_\alpha\, .
$$
From this it follows that 
$$
P_{\alpha} P_{\beta} h^{1/2}((\alpha,j),(\beta,j)) P_\alpha P_\beta
  = \frac{1}{(2\Spin)^2} \sum_{k,l = 1}^{2\Spin} P_{\alpha} P_{\beta} h^{1/2}((\alpha,k),(\beta,l)) P_\alpha P_\beta\, ,
$$
and finally that 
$$
P_\Lambda \widetilde{\widetilde{H}}_{\widetilde{\Lambda}} P_\Lambda
  = \frac{1}{2\Spin} P_\Lambda \widetilde{H}_{\widetilde{\Lambda}} P_\Lambda\, .
$$
Then we can rewrite the spectral gap formula once more
\begin{equation}
\label{sp gap first}
\gamma(\Lambda,\Spin,M) = 2 \Spin 
  \inf_{\substack{\psi \in \Hil(\widetilde{\Lambda},M) \setminus \ker P_\Lambda \\
  \psi \perp \GS(\widetilde{\Lambda},M)}} 
  \frac{\ip{P_\Lambda \psi}{\widetilde{\widetilde{H}}_{\widetilde{\Lambda}} P_\Lambda \psi}}
  {\ip{P_\Lambda \psi}{P_\Lambda \psi}}\, .
\end{equation}
This is useful because we have a spectral gap for $\widetilde{\widetilde{H}}_{\widetilde{\Lambda}}$.
Also note that it is now trivial that $\GS(\widetilde{\Lambda},M) \subset \Hil_0(\widetilde{\Lambda},M)$.
We define 
$$
\Hil_{0,\perp}(\widetilde{\Lambda},M) = \Hil_{0}(\widetilde{\Lambda},M) \cap \GS(\widetilde{\Lambda},M)^{\perp}\, .
$$
Then
$$
\Hil(\widetilde{\Lambda},M) = \GS(\widetilde{\Lambda},M) \oplus
  \Hil_{0,\perp}(\widetilde{\Lambda},M) \oplus \Hil_{\textrm{exc}}(\widetilde{\Lambda},M)\, .
$$
We now state the main lemma of this section, which is the key to our theorem
\begin{lemma}
\label{Key lemma}
\label{Spin ladder bound}
If $\psi \in \Hil(\widetilde{\Lambda},M)$ and $\psi \perp \GS(\widetilde{\Lambda},M)$,
then for some $\psi' \in \Hil_{0,\perp}(\widetilde{\Lambda},M)$ and 
$\psi'' \in \Hil_{\textrm{exc}}(\widetilde{\Lambda},M)$, we have
$$
P_\Lambda \psi = \psi' + \psi''\, .
$$
Moreover,
$$
\frac{\ip{P_\Lambda \psi}{\widetilde{\widetilde{H}}_{\widetilde{\Lambda}} P_\Lambda \psi}}
{\ip{P_\Lambda \psi}{P_\Lambda \psi}}
  \geq (1 - \Delta^{-1})\left(1 - \frac{\ip{\psi'}{P_\Lambda \psi'}}{\ip{\psi'}{\psi'}}\right)
$$
where the ratio is interpreted as zero if $\psi'=0$.
Hence
$$
\gamma(\Lambda,\Spin,M) \geq 2 \Spin (1 - \Delta^{-1}) \left(1 - 
  \sup_{\psi \in \Hil_{0,\perp}(\widetilde{\Lambda},M)} \frac{\ip{\psi}{P_\Lambda \psi}}{\ip{\psi}{\psi}}\right)\, .
$$
\end{lemma}

\begin{proof}
First note that $P_\Lambda \GS(\widetilde{\Lambda},M) = \GS(\widetilde{\Lambda},M) $,
so if $\psi \perp \GS(\widetilde{\Lambda},M)$
then $P_\Lambda \psi \perp \GS(\widetilde{\Lambda},M)$, which proves that 
$P_\Lambda \psi \in \Hil_{0,\perp}(\widetilde{\Lambda},M) \oplus \Hil_{\textrm{exc}}(\widetilde{\Lambda},M)$.
Now suppose $P_\Lambda \psi = \psi' + \psi''$.
Then $\widetilde{\widetilde{H}}_{\widetilde{\Lambda}} \psi ' = 0$.
Hence, by  Theorem \ref{Spin-1/2 spectral gap}
$$
\ip{P_\Lambda \psi}{\widetilde{\widetilde{H}}_{\widetilde{\Lambda}} P_\Lambda \psi}
  = \ip{\psi''}{\widetilde{\widetilde{H}}_{\widetilde{\Lambda}} \psi''} 
  \geq (1 - \Delta^{-1}) \ip{\psi''}{\psi''} \, .
$$
So
$$
\frac{\ip{P_\Lambda \psi}{\widetilde{\widetilde{H}}_{\widetilde{\Lambda}} P_\Lambda \psi}}
{\ip{P_\Lambda \psi}{P_\Lambda \psi}}
  \geq (1 - \Delta^{-1})\left(1 - \frac{\|\psi'\|^2}{\|P_\Lambda \psi\|^2}\right)\, .
$$
By Cauchy-Schwarz and the fact that $\ip{\psi'}{\psi''} = 0$,
we have
$$
\ip{\psi'}{\psi'} = \ip{\psi'}{P_\Lambda \psi}
  = \ip{P_\Lambda \psi'}{P_\Lambda \psi}
  \leq \|P_\Lambda \psi'\| \cdot \|P_\Lambda \psi\|\ .
$$
I.e.\ 
$$
\frac{\|\psi'\|}{\|P_\Lambda \psi\|} \leq \frac{\|P_\Lambda \psi'\|}{\|\psi'\|}\, .
$$
\end{proof}
Thus we may define 
$$
\delta(\Lambda,\Spin,M) = \sup_{\psi \in \Hil_{0,\perp}(\widetilde{\Lambda},M)} 
  \frac{\ip{\psi}{P_\Lambda \psi}}{\ip{\psi}{\psi}}\, .
$$
The positivity of a gap $\gamma(\Lambda,\Spin,M)$ is then equivalent to $\delta(\Lambda,\Spin,M) < 1$.
We will numerically estimate $\delta(\Lambda,\Spin,M)$ in
Section 6, directly from its definition.
For now, we use the lemma to prove the main theorem.


\section{Proof of Theorem \ref{Our theorem}}

We begin this section with the trivial part of the theorem, namely the calculation of
the spectral gap above the state $\omega^{\uparrow}_\Spin$ which is the translation invariant
all up spin state.
This is a well-known result, but we include it for completeness.
To prove that there is a spectral gap, we have to prove that there is a number $\gamma_\Spin >0$
such that for any local observable 
$$
\omega^{\uparrow}_{\Spin}(\delta(X)^*\delta(\delta(X)) - \gamma_{\Spin}\delta(X^*)\delta(X)) \geq 0\, ,
$$
where $\delta(X) = \lim_{\Lambda \nearrow \infty} [H^{\Spin}_\Lambda,X]$.
If $X \in \Obs_\Lambda$ is local, we can take 
$\delta(\delta(X)) = [H^{\Spin}_{\Lambda+[-2,2]},[H^{\Spin}_{\Lambda+[-1,1]},X]]$
and so on.
We observe that the boundary terms of $H^{\Spin}$ are equal to zero, i.e.\
$$
\lim_{L \to \infty} \omega^\uparrow_{\Spin}
(\delta(X)^* (S_{-L}^3 - S_{L}^3) \delta(X)) = 0\, .
$$
So we may rewrite the Hamiltonian in the GNS space of $\omega^{\uparrow}_{\Spin}$
\begin{align*}
H^{\Spin} &= \Delta^{-1} H^{\Spin}_{\textrm{iso}} + (1 - \Delta^{-1}) H^{\Spin}_{\textrm{Ising}}\, ,\\
H^{\Spin}_{\textrm{iso}} &= \sum_{x=-\infty}^{\infty} (\Spin^2 - \boldsymbol{S}_x \cdot \boldsymbol{S}_{x+1})\, ,\\
H^{\Spin}_{\textrm{Ising}} &= \sum_{x=-\infty}^{\infty} (\Spin^2 - S_x^3 S_{x+1}^3) \, .
\end{align*}
Clearly $H^{\Spin}_{\textrm{iso}} \geq 0$, so $\gamma_{\Spin}$ is bounded below by
the spectral gap of $H^{\Spin}_{\textrm{Ising}}$.
It is important that $\omega^{\uparrow}_{\Spin}$ is a ground state of both
$H^{\Spin}_{\textrm{iso}}$ and $H^{\Spin}_{\textrm{Ising}}$.
It is easy to see
that the first excitations for $H^{\Spin}_{\textrm{Ising}}$ are the
one magnon states.
The one magnon states are those obtained from $\omega^{\uparrow}_{\Spin}$ by conjugating with observables
of the form
$$
X = \sum_{x \in \Ir} \frac{c_x}{\sqrt{2 \Spin}} S_x^-
$$
where $\{c_x\}$ is any complex, square-summable sequence.
Then one observes that
$\omega^{\uparrow}_{\Spin}(X) = 0$,
$\omega^{\uparrow}_{\Spin}(X^* X) = 1$, and
$\omega^{\uparrow}_{\Spin}(X^* H^{\Spin}_{\textrm{Ising}} X) = 2 \Spin$.
We claim that it is easy to see that among all quasilocal perturbations, satisfying
the first two equalities, these minimimize
the Ising energy.
So the spectral gap of $H^{\Spin}_{\textrm{Ising}}$ is $2\Spin$, and the spectral
gap of $H^{\Spin}$ is at least $(1 - \Delta^{-1}) 2 \Spin$.
It is also very well known that the Heisenberg model acts as the discrete Laplacian
on the space of one magnon states.
We can see this since
$$
[H^{\Spin}_{\textrm{isotropic}},X] 
  = \sum_{x \in \Ir} \frac{(c_x - c_{x-1}) S_{x-1}^3  + (c_x - c_{x+1}) S_{x+1}^3}{\sqrt{2 \Spin}} S_x^-\, ,
$$
which implies
$$
\omega^{\uparrow}_{\Spin}(X^* H^{\Spin}_{\textrm{isotropic}} X)
  = \Spin \sum_{x=-\infty}^{\infty} \overline{c_x} (2 c_x - c_{x+1} - c_{x-1})\, .
$$
We can choose a sequence of one magnon excitations
$$
X_L = \frac{1}{\sqrt{L}} \sum_{x=1}^L \frac{1}{\sqrt{2\Spin}} S_x^-
$$
such that $\omega^{\uparrow}_{\Spin}(X_L^* H^{\Spin}_{\textrm{isotropic}} X_L) = 2 \Spin/L$.
Therefore,
$$
2 \Spin (1 - \Delta^{-1}) \leq \gamma_{\Spin}
  \leq \omega^{\uparrow}_{\Spin}(X_L^* H^{\Spin} X_L)
  = 2 \Spin (1 - \Delta^{-1}) + 2 \Spin \Delta^{-1} L^{-1}\, ,
$$
for all $L$, which shows that $\gamma_{\Spin} = 2 \Spin (1 - \Delta^{-1})$.

The other trivial facts in the theorem are that for the gaps above the kink
$\gamma(\Spin,M) = \gamma(\Spin,M+2\Spin)$, which follows by translational symmetry,
and $\gamma(\Spin,M) = \gamma(\Spin,-M)$, which follows by spin-flip reflection symmetry.

We now begin the proof of the nontrivial parts of the theorem.
Fix $\Spin \in \frac{1}{2} \Nl$ and $\Delta^{-1}$ in the range $0<\Delta^{-1}<1$.
\begin{definition}
We say that a sequence of triples $(L_k,M_k,\psi_k)$ satisfies hypothesis (H1) if
for all $k \in \Nl$ the following holds: $L_k \in \Nl_{\geq 2}$, 
$M_k \in [-\Spin L_k,\Spin L_k]$, $\psi_k$ in $\Hil_{0,\perp}([1,L_k]\times[1,2\Spin],M_k)$,
and $\|\psi_k\|^2 = 1$.
We note that the most important part of (H1) is that $\psi_k \in \Hil_{0,\perp}$.
In particular, this means that $P_{[1,L_k]} \psi_k \neq \psi_k$.
We say that a sequence of triples $(L_k,M_k,\psi_k)$ satisfies hypothesis (H2) if
additionally 
$$
\lim_{k \to \infty} \|P_{[1,L_k]} \psi_k\| = 1\, .
$$
\end{definition} 
The main component of our proof is the following
\begin{proposition}
\label{Main proposition}
\label{Finite proposition}
No sequence satisfies both hypotheses (H1) and (H2).
\end{proposition}
We observe of both (H1) and (H2) that if any sequence $(L_k,M_k,\psi_k)$ satisfies (H1)
or (H2), then every subsequence does as well.
Using this fact and the previous proposition, one can deduce that there
is a constant $\delta > 0$, depending on $\Delta^{-1}$ and $\Spin$, such that
for any sequence $(L_k,M_k,\psi_k)$ satisfying (H1), one has
$$
\limsup_{k \to \infty} \|P_{[1,L_k]} \psi_k\| \leq 1 - \delta\, .
$$
Hence by Lemma \ref{Key lemma} and the discussion following it,
$$
\inf_{L \in \Nl_{\geq 2}} \inf_{-\Spin L \leq M \leq \Spin L} \gamma([1,L],\Spin,M)
  \geq 2 \Spin (1 - \Delta^{-1}) \delta\, .
$$
All the finite volume spectral gaps have a uniform lower bound.
Since the kink ground states are frustration free, this gives
the following corollary, which is a reformulation of our main theorem
\begin{corollary}
\label{main prop corollary}
For $0<\Delta^{-1}<1$ and any $\Spin \in \frac{1}{2} \Nl$, there is a nonvanishing spectral
gap above all of the infinite volume kink states.
\end{corollary}

\begin{proof}
(\textbf{of Corollary \ref{main prop corollary} given Proposition 
\ref{Main proposition}})
Let $\omega^{\downarrow\uparrow}_{\Spin,M}$ be the ground state of the kink.
To prove that there is a spectral gap, we have to prove that there is a number $\gamma(\Spin,M)>0$
such that for any local observable 
$$
\omega^{\downarrow\uparrow}_{\Spin,M}(\delta(X)^*\delta(\delta(X)) - \gamma(\Spin,M)\delta(X^*)\delta(X)) \geq 0\, ,
$$
where $\delta(X) = \lim_{\Lambda \nearrow \infty} [H_\Lambda,X]$.
On the other hand, by Lemma \ref{kink state convergence}, 
for any local observable $X$
$$
\omega^{\downarrow\uparrow}_{\Spin,M}(X) 
  = \lim_{L \to \infty} \frac{\ip{\Psi_0(\Lambda_L,\Spin,M)}{X \Psi_0(\Lambda_L,\Spin,M)}}
  {\|\Psi_0(\Lambda_L,\Spin,M)\|^2}\, ,
$$
where $\Lambda_L = [-L+1,L]$.
If $X$ is a local observable with support in $\Lambda$ then one can take 
$\delta(\delta(X)) = [H_{\Lambda_L},[H_{\Lambda_L},X]]$, for any $\Lambda_L \supset \Lambda + [-2,2]$.
Also for $M' \in [-\Spin|\Lambda_L|,\Spin|\Lambda_L|]$,
$$
\ip{\Psi_0(\Lambda_L,\Spin,M')}{(H_{\Lambda_L} X - X H_{\Lambda_L})\Psi_0(\Lambda_L,\Spin,M)}
  = 0
$$
because $\Psi_0(\Lambda_L,\Spin,M')$ and $\Psi_0(\Lambda_L,\Spin,M)$ have the same energy.
Hence $\delta(X) \Psi_0(\Lambda_L,\Spin,M)$ is orthogonal to all ground states.
Thus, we see that in this case the infinite volume gap corresponds to the naive guess, i.e.\ the
liminf of all finite volume gaps.
To be more explicit,
let $\gamma(M,\Spin)$ be a positive lower bound on all the finite volume gaps $\gamma([-L+1,L],\Spin,M)$,
which exists by Proposition \ref{Main proposition}. 
Define $\omega_L$ and $\omega'_L$ to be the states on $\Obs_{\Lambda_L}$ 
corresponding to $\Psi_0(\Lambda_L,\Spin,M)$ and $\delta(X) \Psi_0(\Lambda_L,\Spin,M)$, respectively.
Then
\begin{align*}
&\omega^{\downarrow\uparrow}_{\Spin,M}(\delta(X)^* \delta(\delta(X)) - \gamma(\Spin,M)\delta(X^*)\delta(X)) \\
  &\qquad = \lim_{L \to \infty} \omega_L(\delta(X)^* \delta(\delta(X)) 
- \gamma(\Spin,M)\delta(X^*)\delta(X)) \\
  &\qquad = \lim_{L \to \infty} \omega_L'(H_{\Lambda_L} - \gamma(\Spin,M))
  \omega_L(\delta(X)^* \delta(X)) \\
  &\qquad \geq \lim_{L \to \infty}
  (\gamma(\Lambda_L,\Spin,M) - \gamma(\Spin,M)) \frac{\|\delta(X) \Psi_0(\Lambda_L,\Spin,M)\|^2}
  {\|\Psi_0(\Lambda_L,\Spin,M)\|^2}\\
  &\qquad \geq 0\, .
\end{align*}
\end{proof}

We will show that (H1) and (H2) are incompatible
in a series of lemmas.
\begin{lemma}
\label{finite finite}
If $(L_k,M_k,\psi_k)$ satisfies (H1) and (H2), then $L_k \to \infty$.
\end{lemma}
\begin{proof}
{\textbf (of Lemma \ref{finite finite})}
Suppose not. Then there is some sequence satisfying (H1) and (H2) and such that
$L_k = L<\infty$ for all $k$.
But $\Hil_0([1,L],\Spin) = \bigoplus_{M=-\Spin L}^{\Spin L} \Hil_0([1,L],\Spin,M)$ is a finite dimensional space.
The finite matrix obtained by restricting and projecting $P_{[1,L]}$ to this space is Hermitian, 
and its largest eigenvalue is 1.
Moreover, the eigenspace corresponding to 1 is $\GS([1,L],\Spin) = \bigoplus_{M=-\Spin L}^{\Spin L} \GS([1,L],\Spin,M)$.
All the vectors $\psi_k$ are orthogonal to $\GS([1,L],\Spin)$ by hypothesis (H1).
Since finite matrices have discrete spectra, this contradicts hypothesis (H2).
\end{proof}

\begin{lemma}
\label{finite}
If $(L_k,M_k,\psi_k)$ satisfies (H1) and (H2) then $\Spin L_k - |M_k| \to \infty$.
\end{lemma}

Before giving the proof of this easy lemma, we need to define some new notation.
\begin{definition}
For $L \in \Nl$, let $\mathcal{M}(L,\Spin,M)$ be 
the set of all vectors $\boldsymbol{m}=(m_1,\dots,m_{2\Spin}) \in [-L/2,L/2]^{2\Spin}$
whose sum is $M$, i.e.\ $\sum_{j} m_j = M$.
For each $\boldsymbol{m} \in \mathcal{M}(L,\Spin,M)$, define
$$
\Psi_0([1,L] \times [1,2\Spin],\boldsymbol{m})
  = \bigotimes_{j=1}^{2\Spin} \Psi_0([1,L]\times\{j\},1/2,m_j)\, .
$$
Related to this, let $\mathcal{N}(\Spin,N)$ be the set all vectors 
$\boldsymbol{n} = (n_1,\dots,n_{2\Spin}) \in \Nl^{2\Spin}$
satisfying $\sum_{j} n_j = N$.
Let $\boldsymbol{e} = (1,1,\dots,1) \in \Nl^{2\Spin}$.
Recall a previous definition
$$
\Psi_0'([1,L],1/2,n) = \sum_{1\leq x_1<x_2<\dots<x_n\leq L} q^{x_1+\dots+x_n}
  S_{x_1}^- \cdots S_{x_n}^- \Omega_\Nl\, .
$$
Define 
$$
\Psi_0'([1,L]\times[1,2\Spin],\boldsymbol{n})
  = \bigotimes_{j=1}^{2\Spin} \Psi_0'([1,L]\times\{j\},1/2,n_j)
$$
Let $\Hil(\Nl\times[1,2\Spin],\textrm{up})$ be the Guichardet Hilbert space 
$$
\bigotimes_{(x,j)\in \Nl[1,2\Spin]} (\Cx^2_{(x,j)},\ket{+1/2}_{(x,j)})\, .
$$
Define $\mathcal{D}([1,L]\times[1,2\Spin],N)$ to be the projection on 
$\Hil(\Nl\times[1,2\Spin],\textrm{up})$
which projects onto vectors $\psi$ such that 
$$
\sum_{(x,j) \in [1,L]\times [1,2\Spin]} (\frac{1}{2} - S_{(x,j)}^3) \psi = N \psi\, ,
$$
and $S^3_{(x,j)} \psi = \frac{1}{2} \psi$ for $(x,j) \in (\Nl \setminus [1,L]) \times [1,2\Spin]$.
Finally, let 
$$
\Psi_0'([1,L]\times[1,2\Spin],N)
$$ 
be the unique 
normalized ground state of $\widetilde{H}_{[1,L]\times[1,2\Spin]}$
in the range of $\mathcal{D}([1,L]\times[1,2\Spin],N)$.
Specify the phase to that $\Psi_0'([1,L]\times[1,2\Spin],N)$ has real coefficients
in the Ising basis.
This ground state exists and is unique (and has real coefficients)
since $\widetilde{H}_{[1,L]\times[1,2\Spin]}$
acting on the range $\mathcal{D}([1,L]\times[1,2\Spin],N)$ is unitarily equivalent to the 
finite matrix $\widetilde{H}_{[1,L]\times[1,2\Spin]}$ acting on 
$\Hil([1,L]\times[1,2\Spin],M)$.
(We apologize for the abuse of notation : the same notation is used for the operator acting
on two different, but isomorphic, Hilbert spaces.)
\end{definition}

The following are easy and useful observations.
The set 
$$
\{\Psi_0([1,L]\times[1,2\Spin],\boldsymbol{m}) : \boldsymbol{m} \in \mathcal{M}(L,\Spin,M)\}
$$
is an orthogonal basis for $\Hil_0([1,L]\times[1,2\Spin],M)$.
The set of indices $\mathcal{N}(\Spin,n)$ is finite, while if one defined the analogue
of $\mathcal{M}(L,\Spin,M)$ replacing $L$ by $\Nl$ it would not be finite.
There is a simple translation between the two vectors defined above:
$$
\frac{\Psi_0'([1,L]\times[1,2\Spin],\boldsymbol{n})}
{\|\Psi_0'([1,L]\times[1,2\Spin],\boldsymbol{n})\|}
  = \frac{i_{[1,L],\Ir} \Psi_0([1,L]\times[1,2\Spin],\frac{1}{2} L \boldsymbol{e} - \boldsymbol{n})}
{\|\Psi_0([1,L]\times[1,2\Spin],\frac{1}{2} L \boldsymbol{e} - \boldsymbol{n})\|}\, .
$$
By Lemma \ref{kink state convergence}, the following strong limit exists
$$
\Psi_0'(\Nl\times[1,2\Spin],\boldsymbol{n}) = \lim_{L\to \infty} \Psi_0'([1,L]\times[1,2\Spin],\boldsymbol{n})\, .
$$
One has the following simple formula for the action of $\mathcal{D}([1,L]\times[1,2\Spin],N)$ on 
$\Hil_0([1,L']\times[1,2\Spin],M)$:
$$
\mathcal{D}([1,L]\times[1,2\Spin],N) \Psi_0'([1,L']\times[1,2\Spin],\boldsymbol{n})
  = \Psi_0'([1,L'\wedge L]\times[1,2\Spin],\boldsymbol{n})\, .
$$
By Lemma \ref{kink state convergence}, again, the following limit 
holds for all $\boldsymbol{n} \in \mathcal{N}(\Spin,N)$:
\begin{equation}
\label{nec1}
\lim_{L \to \infty} \lim_{L' \to \infty}
  \frac{\|\mathcal{D}([1,L]\times[1,2\Spin],N) \Psi_0'([1,L']\times[1,2\Spin],\boldsymbol{n})\|}
  {\|\Psi_0'([1,L']\times[1,2\Spin],\boldsymbol{n})\|}
  = 1\, .
\end{equation}
By Lemma \ref{kink state convergence}, again, the following strong limit exists
\begin{equation}
\label{nec3}
\Psi_0'(\Nl \times[1,2\Spin],N) = \lim_{L\to \infty} \Psi_0'([1,L]\times[1,2\Spin],N)\, .
\end{equation}
We note for the reader that for large enough $L$
$$
\Psi_0'([1,L] \times[1,2\Spin],N) \propto \sum_{\boldsymbol{n} \in \mathcal{N}(\Spin,N)} 
  \Psi_0'([1,L]\times[1,2\Spin],\boldsymbol{n})\, ,
$$
and that
$$
\Psi_0'(\Nl \times[1,2\Spin],N) \propto \sum_{\boldsymbol{n} \in \mathcal{N}(\Spin,N)} 
  \Psi_0'(\Nl\times[1,2\Spin],\boldsymbol{n})\, .
$$
The proportionality constants are necessary because both $\Psi_0'([1,L]\times[1,2\Spin],N)$
and $\Psi_0'(\Nl\times[1,2\Spin],N)$ are chosen to be normalized.

\begin{proof} (\textbf{Lemma \ref{finite}})
We will verify that $\Spin L_k - M_k \to \infty$; the fact that $\Spin L_k + M_k \to \infty$
then follows by spin-flip/reflection symmetry.
If it fails for some sequence, then some subsequence, also satisfying 
(H1) and (H2), has the property that $\Spin L_{l_k} - M_{l_k} = M$ is a finite
constant.
We assume that this subsequence was taken at the beginning to avoid double subscripts.
Since each $\psi_k \in \Hil_{0,\perp}([1,L_k]\times[1,2\Spin],M)$, we can write
$$
\psi_k = \sum_{\boldsymbol{n} \in \mathcal{N}([1,L_k],N)} c_k(\boldsymbol{n})
  \frac{\Psi_0([1,L_k]\times[1,2\Spin],\frac{1}{2}L_k \boldsymbol{e} - \boldsymbol{n})}
  {\|\Psi_0([1,L_k]\times[1,2\Spin],\frac{1}{2}L_k \boldsymbol{e} - \boldsymbol{n})\|}\, .
$$
So
$$
\psi_k' := 
  i_{[1,L_k],\Nl} \psi_k 
  = \sum_{\boldsymbol{n} \in \mathcal{N}([1,L_k],N)} c_k(\boldsymbol{n})
  \frac{\Psi_0'([1,L_k]\times[1,2\Spin],\boldsymbol{n})}
  {\|\Psi_0'([1,L_k]\times[1,2\Spin],\boldsymbol{n})\|}\, .
$$
We know $L_k \to \infty$ by the last lemma.
So $\Psi_0'([1,L_k]\times[1,2\Spin],\boldsymbol{n}) \to \Psi_0'(\Nl\times[1,2\Spin],\boldsymbol{n})$
as $k \to \infty$ for each $\boldsymbol{n}$.
Furthermore, $\{c_k(\boldsymbol{n}) : \boldsymbol{n} \in \mathcal{N}(\Spin,N)\}$
is a unit vector in the finite-dimensional space $\Cx^{|\mathcal{N}(\Spin,N)|}$, for each $k$.
So there is a subsequence with $c_{l_k}(\boldsymbol{n}) \to c(\boldsymbol{n})$ for each $\boldsymbol{n}$.
Again, we assume this subsequence was chosen at the beginning to avoid double subscripts.
Thus, $\psi_{k}'$ converges to the vector
$$
\psi = \sum_{\boldsymbol{n} \in \mathcal{N}(\Spin,N)} c(\boldsymbol{n})
  \Psi_0'(\Nl\times[1,2\Spin],\boldsymbol{n})\, .
$$

For $L<\infty$ and $k$ large enough, $P_{[1,L]} \geq P_{[1,L_{l_k}]}$.
Hence, by (H2)
\begin{align*}
\ip{\psi}{P_{[1,L]} \psi} 
  = \lim_{k\to \infty} \ip{\psi'_{k}}{P_{[1,L]} \psi_{k}'}
  \geq \lim_{k\to \infty} \ip{\psi'_{k}}{P_{[1,L_{k}]} \psi'_{k}}
  = 1
\end{align*}
for all $L$.
Also, $\widetilde{\widetilde{H}}_{[1,L_{k}]\times[1,2\Spin]} \geq 
\widetilde{\widetilde{H}}_{[1,L]\times[1,2\Spin]}$,
so $\widetilde{\widetilde{H}}_{[1,L]\times[1,2\Spin]} \psi = 0$.
Recall
$$
P_{[1,L]} \widetilde{\widetilde{H}}_{[1,L]\times[1,2\Spin]} P_{[1,L]}
  = \frac{1}{2\Spin} P_{[1,L]} \widetilde{H}_{[1,L]\times[1,2\Spin]} P_{[1,L]}\, .
$$
So
\begin{equation}
\label{ground state equation}
\widetilde{H}_{[1,L]\times[1,2\Spin]} \psi = 0
\end{equation}
for every finite $L$.

Notice,
\begin{equation}
\label{nec2}
\lim_{L \to \infty} \|\mathcal{D}([1,L]\times[1,2\Spin],N) \psi\|
  = \lim_{L\to\infty} \lim_{k \to \infty} \|\mathcal{D}([1,L]\times[1,2\Spin],N) \psi_k\|
  = 1
\end{equation}
by \eq{nec1}.
Together, \eq{ground state equation} and \eq{nec2} imply that
$$
\lim_{L \to \infty} |\ip{\psi}{\Psi_0'([1,L]\times[1,2\Spin],N)}| = 1\, ,
$$
i.e.\ that $\psi = e^{i\phi} \Psi_0'(\Nl\times[1,2\Spin],N)$.
Since $\psi'_k \to \psi$,
\begin{equation}
\lim_{k\to\infty} \frac{|\ip{\psi'_k}{\Psi_0'(\Nl\times[1,2\Spin],N)}|}
  {\|\psi'_k\|} = 1\, .
\label{is1}\end{equation}
On the other hand, we know $\ip{\psi'_k}{\Psi_0'([1,L_k]\times[1,2\Spin],N)} = 0$,
because $\psi_k \perp \GS([1,L_k]\times[1,2\Spin],N)$.
So, by \eq{nec3}, this implies
\begin{equation}
\lim_{k \to \infty} \frac{|\ip{\psi'_k}{\Psi_0'(\Nl\times[1,2\Spin],N)}|}
  {\|\psi'_k\|} = 0\, .
\label{is0}\end{equation}
Clearly, \eq{is1} and \eq{is0} are incompatible and we have a contradiction.
\end{proof}

\begin{proof}
(\textbf{Proposition \ref{Finite proposition}})

Let $\mathcal{M}(\Spin,M)$ be the set of all $(m_1,\dots,m_{2\Spin})$ with
$\sum_j m_j = M$.
Define for any $m \in \Ir$,
\begin{align*}
\Psi_0'([-a,b],1/2,m) &= \sum_n 
  \sum_{-a\leq \alpha_1<\dots<\alpha_n\leq 0<\beta_1 < \dots < \beta_{m+n}\leq b} \\
  &q^{\beta_1 + \dots \beta_{m+n} - \alpha_1 - \dots - \alpha_n} S_{\alpha_1}^+ \cdots S_{\alpha_n}^+
  S_{\beta_1}^- \cdots S_{\beta_{m+n}}^- \Omega_\Ir\, .
\end{align*}
Let
$$
\Psi_0'([-a,b]\times[1,2\Spin],\boldsymbol{m})
  = \bigotimes_{j=1}^{2\Spin} \Psi_0'([-a,b]\times\{j\},1/2,m_j)\, .
$$
Suppose that $(L_k,M_k,\psi_k)$ is a sequence satisfying (H1) and (H2).
By Lemmas \ref{finite finite} and  \ref{finite}
we may assume that $L_k \to \infty$ and $\Spin L_k - |M_k| \to \infty$. 
We may choose a subsequence so that $\Spin L_k - M_k = 2 \Spin r_k - N$
for all $k$, where $r_k \in \Nl_{\geq0}$ is arbitrary and $N \in [0,2\Spin-1]$ is fixed.
Let
$$
\psi_k' = T^{-r_k} i_{[1,L_k],\Ir} \psi_k\, .
$$
let $\Lambda_k = [1-r_k,L_k-r_k]$.
Then we can write
$$
\psi_k' = \sum_{\boldsymbol{m} \in \mathcal{M}(\Spin,N)} C_k(\boldsymbol{m})
  \frac{\Psi_0'(\Lambda_k \times[1,2\Spin],\boldsymbol{m})}
  {\|\Psi_0'(\Lambda_k \times[1,2\Spin],\boldsymbol{m})\|}\, .
$$
Now, for each $k$, $\{C_k(\boldsymbol{m}) : \boldsymbol{m} \in \mathcal{M}(\Spin,N)\}$
is a normalized $l^2$ sequence, instead of a finite dimensional vector.
So we only know that a weakly convergent subsequence exists, not a strongly convergent one.
We need some kind of tightness result, which is given in the  following lemma.
\begin{lemma}
\label{Technical lemma}
Let $\mathcal{M}_R(\Spin,N)$ be those 
$\boldsymbol{m} \in \mathcal{M}(\Spin,N)$
such that all components lie in the range $[-R+1,R]$.
Then we have
$$
\lim_{R \to \infty} \liminf_{k \to \infty} 
  \sum_{\boldsymbol{m} \in \mathcal{M}_R(\Spin,N)} 
  |C_k(\boldsymbol{m})|^2 = 1\, .
$$
\end{lemma}

We give the proof of this technical but important lemma at the end of the section.
First, we see how Proposition \ref{Finite proposition} follows.

\begin{lemma}
\label{Second tech lemma}
Define $\mathcal{F}_R$ to be the projection onto those vectors with all down spins at sites $(\alpha,j)$
when $\alpha\leq -R$, 
all up spins at sites $(\alpha,j)$ with $\alpha\geq R+1$, and $S^3_{[-R+1,R]\times[1,2\Spin]}$ equal to 
$N$.
We have
$$
\lim_{R \to \infty} \liminf_{k \to \infty} \|\mathcal{F}_R \psi_k'\| = 1\, .
$$
\end{lemma}

\begin{proof}
By lemma \ref{Technical lemma}, for any $\epsilon>0$, we can choose $R$ large enough that
$$
\liminf_{k \to \infty} \sum_{\boldsymbol{m} \in \mathcal{M}_R(\Spin,M)} 
  |C_k(\boldsymbol{m})|^2 > 1-\epsilon\, .
$$
By Lemma \ref{kink state convergence}, 
the following strong limit exists
$$
\Psi_0'(\Ir\times[1,2\Spin],\boldsymbol{m}) =
\lim_{k \to \infty} \Psi_0'(\Lambda_k\times [1,2\Spin],\boldsymbol{m})\, .
$$
Furthermore,
$$
\lim_{R_1 \to \infty} \frac{\|\mathcal{F}_{R_1} \Psi_0'(\Ir\times[1,2\Spin],\boldsymbol{m})\|}
  {\|\Psi_0'(\Ir\times[1,2\Spin],\boldsymbol{m})\|} = 1\, .
$$
Thus
$$
\liminf_{R_1 \to \infty} \|\mathcal{F}_{R_1} \psi'_k\|^2 > 1 - \epsilon\, .
$$
Since $\epsilon$ was arbitrary, we are done.
\end{proof}

We know that $\widetilde{\widetilde{H}}_{\widetilde{\Lambda}_k} \psi'_k = 0$
and $\lim_k \|P_{\Lambda_k} \psi'_k\| = 1$.
As before, these two facts are sufficient to guarantee that for any finite $\Lambda \subset \Ir$,
$$
\lim_{k \to \infty} \widetilde{H}_{\widetilde{\Lambda}} \psi_k' = 0\, .
$$
In other words, letting $G_\Lambda$ be the projection onto the ground state space of 
$\widetilde{H}_{\widetilde \Lambda}$, that 
\begin{equation}
\label{third tech}
\lim_{k \to \infty} \|G_\Lambda \psi_k'\| = 1\, .
\end{equation}
We note that for $\Lambda = [-R+1,R]$, the projections $\mathcal{F}_R$ and $G_\Lambda$
commute, and in fact their product is the projection onto the normalized ground state vector
$$
\Psi_0'([-R+1,R]\times[1,2\Spin],N)\, .
$$
By Lemma \ref{Second tech lemma} and equation \eq{third tech}
we see that
$$
\lim_{R \to \infty} \liminf_{k \to \infty} 
  |\ip{\Psi_0'([-R+1,R]\times[1,2\Spin],N)}{\psi_k'}|^2 = 1\, .
$$
Since $\Psi_0'([-R+1,R]\times[1,2\Spin],N)$ converges in norm to 
$\Psi_0'(\Ir\times[1,2\Spin],N)$ as $R \to \infty$, we have
\begin{equation}
\label{final contradiction}
\liminf_{k \to \infty} 
  |\ip{\Psi_0'(\Ir\times[1,2\Spin],N)}{\psi'_k}|^2 = 1\, .
\end{equation}

Now comes the contradiction.
We know, by virtue of the fact that $\psi'_k \in \widetilde{\Hil}_{0,\perp}([1,L_k]\times[1,2\Spin],M_k)$,
that $\psi_k' \perp \Psi_0'([1,L_k]\times[1,2\Spin],N)$.
But on the other hand, we know that $\Psi_0'([1,L_k]\times[1,2\Spin],N)$
converges strongly to $\Psi_0'(\Ir\times[1,2\Spin],N)$, so
$$
\limsup_{k \to \infty} 
  |\ip{\Psi_0'(\Ir\times[1,2\Spin],N)}{\psi_k'}|^2 = 0\, ,
$$
clearly contradicting \eq{final contradiction}.
\end{proof}

\begin{proof}
\textbf{(of Lemma \ref{Technical lemma})}
We define $D_R$ to be the projection onto all those vectors in 
$\Hil(\Ir\times[1,2\Spin],\textrm{kink})$ with at most $2\Spin-1$ down spins shared between
the sites $(-R+1,1),\dots,(-R+1,2\Spin)$, and at least one down spin shared between the
sites $(R,1),\dots,(R,2\Spin)$.
Let
$$
\widetilde{\Psi}_0(\widetilde{\Lambda},\boldsymbol{m})
  = \frac{\Psi_0'(\widetilde{\Lambda},\boldsymbol{m})}
  {\|\Psi_0'(\widetilde{\Lambda},\boldsymbol{m})\|}\, .
$$
Let $\Lambda_R = [-R+1,R]$.
It is clear that if 
$$
\psi = \sum_{\boldsymbol{m} \in \mathcal{M}(\Spin,M)}
  C(\boldsymbol{m}) \widetilde{\Psi}_0(\widetilde{\Lambda}_R,\boldsymbol{m})
$$
then
\begin{align*}
\ip{\psi}{D_R \psi}
  &= \sum_{\boldsymbol{m}}
  |C(\boldsymbol{m})|^2 \|D_R \widetilde{\Psi}_0(\widetilde{\Lambda_R},\boldsymbol{m})\|^2 \\
  &\geq \sum_{\boldsymbol{m} \in \mathcal{M}(\Spin,M) \setminus \mathcal{M}_R(\Spin,M)}
  |C(\boldsymbol{m})|^2 \|D_R \widetilde{\Psi}_0(\widetilde{\Lambda_R},\boldsymbol{m})\|^2 \\
  &\geq \frac{1}{2} \sum_{\boldsymbol{m} \in \mathcal{M}(\Spin,M)
  \setminus \mathcal{M}_R(\Spin,M)}
  |C(\boldsymbol{m})|^2 
\end{align*}
the second line owing to the fact that $D_R$ commutes with 
$S^3_{[-L+1,L]\times\{j\}}$ for each leg $j$ and all $L \in \Nl$.
The last inequality is the result from to the following consideration:
if one of the $m_j$ is greater than $R$ or less than $-R+1$,
then with probability at least $1/2$ one finds the state with at least one down spin at $R$, or 
at least one up spin at $-R+1$, respectively.
Thus
$$
\sum_{\boldsymbol{m} \in \mathcal{M}_R(\Spin,M)}
  |C(\boldsymbol{m})|^2 \geq 1 - 2 \ip{\psi}{D_R \psi}\, .
$$
Now suppose that $\psi$ is a ground state vector of $\widetilde{\widetilde{H}}_{\widetilde{\Lambda}_R}$.
Then, because $D_{R}$ commutes with $S^3_{\widetilde{\Lambda_R}}$
we have
$$
\ip{\psi}{D_R \psi} \leq \max_{M \in [-2 \Spin R,2 \Spin R]} 
  \ip{\Psi_0(\widetilde{\Lambda}_R,M)}{D_R\Psi_0(\widetilde{\Lambda}_R,M)}\, .
$$

We can make the following crude but simple estimate
\begin{equation}
\begin{split}
\max_{M \in [-2 \Spin R,2 \Spin R]} 
 \|D_R \Psi_0(\widetilde{\Lambda}_R,M)\|^2
 \leq \frac{4 \Spin^2 R q^{2R}}{1-4 \Spin^2 R q^{2R}}\, .
\end{split}
\end{equation}

We prove this estimate for $M \leq 0$, and the $M\geq 0$ follows by spin flip/reflection symmetry.
Of course if $M = -2\Spin R$, then the estimate is trivial because $\Psi_0$ has all down spins, so the
expectation with $D_R$ is zero.
Suppose $-2\Spin R<M\leq 0$.
Then there is at least one down spin and at most $2 \Spin R$ down spins.
We write $\mathbb{M}(\widetilde{\Lambda}_R,n)$ for all the classical Ising configurations
on the lattice $\widetilde{\Lambda}_R$ with exactly $n$ down spins.
We write $\mathbb{M}(\widetilde{\Lambda}_R,n,j)$
for all the configurations with the
extra constraint that the number of down spins on the sites $(R,1),\dots,(R,2\Spin)$ is $j$.
The normalized ground state $\Psi_0(\widetilde{\Lambda}_R,2 \Spin R - n)$ is
\begin{align*}
\Psi_0(\widetilde{\Lambda}_R,2\Spin R-n) = Z^{-1}
  \sum_{\{m(\alpha,j)\} \in \mathbb{M}(\widetilde{\Lambda}_R,n)} W(\{m(\alpha,j)\}) \ket{\{m(\alpha,j)\}}\, ,
\end{align*}
where
$$
W(\{m(\alpha,j)\}) = \prod_{(\alpha,j)} q^{-\alpha m(\alpha,j)}
$$
and
$$
Z^2 = 
\sum_{\{m(\alpha,j)\} \in \mathbb{M}(\widetilde{\Lambda}_R,n)} W(\{m(\alpha,j)\})^2
$$
On the other hand, defining $E_{R,j}$ to be the projection onto those states with exactly $j$
down spins on the sites $(R,1),\dots,(R,2\Spin)$,
we have
$$
\|E_{R,j} \Psi_0([-R+1,R]\times [1,2\Spin],2\Spin R -n)\|^2 = \frac{Z_j^2}{Z^2}
$$
where $Z_j$ is the same as $Z$ but with the sum over classical configurations restricted
to $\mathbb{M}(\widetilde{\Lambda}_R,n,j)$.

The inequality comes from recognizing that $Z_{j+1}^2 \leq 4 \Spin^2 R q^{2R} Z_j^2$.
This is a straightforward estimate.
Define a lexicographic order on $\widetilde{\Lambda}_R$
by
$(\alpha_1,j_1) < (\alpha_2,j_2)$ if either $\alpha_1<\alpha_2$ or $\alpha_1=\alpha_2$ and $j_1<j_2$.
Define a map $f : \mathbb{M}(\widetilde{\Lambda}_R,n,j+1) \to \mathbb{M}(\widetilde{\Lambda}_R,n,j)$
where the down spin at the greatest site $(\alpha,j)$ with a down spin, is exchanged for the up
spin at the least site $(\beta,k)$ with an up spin.
Note that since there are at most $2 \Spin R$ down spins, the point $(\beta,k)$ must lie in the
subset $[-R+1,0]\times [1,2\Spin]$, so there are at most $2 \Spin R$ choices of $(\beta,k)$.
Similarly, since $j+1\geq 1$, we know there is at least one down spin in the sites
$(R,1),\dots,(R,2\Spin)$.
So there are at most $2\Spin$ choices for $(\alpha,j) = (R,j)$.
Thus, $\# f^{-1}(\{m(\alpha,j)\}) \leq 4 \Spin^2 R$ for any configuration $\{m(\alpha,j)\}$.
We see that
$$
W(f(\{m(x,j)\})) \geq q^{-R} W(\{m(x,j)\})
$$
because the down spin at site $(\alpha,j)=(R,j)$ has moved at least $R$ units to the left to $(\beta,k)$, 
$\beta \leq 0$.
So
\begin{align*}
Z_{j+1}^2 &= \sum_{\{m(\alpha,j)\} \in \mathbb{M}(\widetilde{\Lambda}_R,n,j+1)} W(\{m(\alpha,j)\})^2 \\
  &\leq q^{2R} \sum_{\{m(\alpha,j)\} \in \mathbb{M}(\widetilde{\Lambda}_R,n,j+_1)} W(f(\{m(\alpha,j)\}))^2 \\
  &\leq 4 \Spin^2 R q^{2R} \sum_{\{m(\alpha,j)\} \in \mathbb{M}(\widetilde{\Lambda}_R,n,j)} W(\{m(\alpha,j)\})^2 \\
  &= 4 \Spin^2 R q^{2R} Z_j^2\, .
\end{align*}
The crude estimate is proved.

Now we know that as $k \to \infty$, the vectors $\psi_k$ come closer and closer to the
ground state space of $\widetilde{H}_{\widetilde{\Lambda}_R}$.
So our estimate implies
$$
\liminf_{k \to \infty} 
\sum_{\boldsymbol{m} \in \mathcal{M}_R(\Spin,M)}
  |C_k(\boldsymbol{m})|^2 \geq 1 - \frac{8 \Spin^2 R q^{2R}}{1 - 4 \Spin^2 R q^{2R}}\, .
$$
Hence
$$
\lim_{R \to \infty} \liminf_{k \to \infty} 
\sum_{\boldsymbol{m} \in \mathcal{M}_R(\Spin,M)}
  |C_k(\boldsymbol{m})|^2 \geq 1 - \lim_{R \to \infty} \frac{8 \Spin^2 R q^{2R}}{1 - 4 \Spin^2 R q^{2R}}
  = 1\, ,
$$
and this concludes the proof.
\end{proof}

\section{Numerical Approximation}
\label{The lower bound matrix}
We now find an explicit representation of
$$ 
\textrm{Proj}(\Hil_0(\widetilde{\Lambda},M))
P_\Lambda
\textrm{Proj}(\Hil_0(\widetilde{\Lambda},M))\, .
$$
From this we numerically calculate $1-\delta(L,\Spin,M)$.
We begin with some definitions.
First of all, we will always have $\Lambda = [1,L]$ in this section,
and hence $\widetilde{\Lambda} = [1,L] \times [1,2\Spin]$.
For $N \in [0,2\Spin L]$, define
$$
\widetilde{P}(\widetilde{\Lambda},N) = \textrm{Proj}(\Hil_0(\widetilde{\Lambda},\Spin L -N))
  P_\Lambda \textrm{Proj}(\Hil_0(\widetilde{\Lambda},\Spin L -N))\, .
$$
Also, define the ``classical Ising configurations'' to be
$$
\mathbb{M}(L,2\Spin,N) = \{A \in [0,1]^{\widetilde{\Lambda}} : \sum_{(x,j) \in \widetilde{\Lambda}} A(x,j) = N\}\, .
$$
These are $\{0,1\}$-matrices with $2\Spin$ rows, $L$ columns, and $N$ ones. 
For any $A \in \mathbb{M}(L,2\Spin,N)$ we define 
$$
\phi_A = \prod_{(x,j) \in \widetilde{\Lambda}} (S_{(x,j)}^-)^{A(x,j)}\, \Omega
$$
where $\Omega = \bigotimes_{(x,j) \in \widetilde{\Lambda}} \ket{1/2}_{(x,j)}$.
Then it is clear that $\Hil(\widetilde{\Lambda},\Spin L - N)$ has an orthonormal basis
$\{\phi_A : A \in \mathbb{M}(L,2\Spin,N)\}$.
For any matrix $A$, we define two vectors $\boldsymbol{r}_A \in \Cx^{2\Spin}$ and 
$\boldsymbol{c}_A \in \Cx^{L}$ by
$$
r_A(j) = \sum_{x =1}^L A(x,j)\, ,\qquad
c_A(x) = \sum_{j=1}^{2\Spin} A(x,j)\, .
$$
Finally we define
\begin{equation}
\label{zero ones matrix}
M_{\boldsymbol{r},\boldsymbol{c}} = \#\{A \in \mathbb{M}(L,2\Spin,N) : \boldsymbol{r}_A = \boldsymbol{r}\, ,\
\boldsymbol{c}_A = \boldsymbol{c}\}\, .
\end{equation}
Note that by its definition $M_{\boldsymbol{r},\boldsymbol{c}}$ is unchanged if one 
permutes the components of $\boldsymbol{r}$ or $\boldsymbol{c}$.
We mention that there is no known formula for $M_{\boldsymbol{r},\boldsymbol{c}}$ although
it has useful characterizations in terms of generating functions. 
(C.f.\ \cite{Sta} \S 7.4 for more details.)

The definitions immediately lead to the following result.
\begin{lemma}
The following are true identities:
\begin{gather}
\Psi_0(\Lambda,\Spin,\frac{1}{2}L \boldsymbol{e} - \boldsymbol{n}) 
  = \sum_{\substack{A \in \mathbb{M}(L,2\Spin,N) \\ \boldsymbol{r}_A = \boldsymbol{n} }}
    \phi_A q^{\boldsymbol{x} \cdot \boldsymbol{c}_A} 
\\
\|\Psi_0(\Lambda,\Spin,\frac{1}{2}L \boldsymbol{e} - \boldsymbol{n})\|^2 
  = \sum_{\substack{\boldsymbol{c} \in [0,2\Spin]^L \\ \sum_x c(x) = N}}
  q^{2 \boldsymbol{x} \cdot \boldsymbol{c}} M_{\boldsymbol{n},\boldsymbol{c}} 
\\
P_\Lambda \phi_A = \prod_{x=1}^L \binom{2\Spin}{c_A(x)} 
  \sum_{\substack{B \in \mathbb{M}(L,2\Spin,N) \\ \boldsymbol{c}_B = \boldsymbol{c}_A}} \phi_B 
\\
\label{symmetrization ip}
\begin{split}
&\ip{\Psi_0(\Lambda,\Spin,\frac{1}{2}L\boldsymbol{e} - \boldsymbol{m})}
  {P_\Lambda \Psi_0(\Lambda,\Spin,\frac{1}{2}L\boldsymbol{e} - \boldsymbol{n})} \\
&\qquad \qquad \qquad \qquad \qquad \qquad 
= \sum_{\substack{\boldsymbol{c} \in [0,2\Spin]^L \\ \sum_x c(x) = N}} M_{\boldsymbol{m},\boldsymbol{c}}
  M_{\boldsymbol{n},\boldsymbol{c}} q^{2\boldsymbol{x}\cdot \boldsymbol{c}} \prod_{x=1}^L \binom{2\Spin}{c(x)}
\end{split}
\end{gather}
\QED
\end{lemma}

We can define an action of $\mathfrak{S}_{2\Spin}$ on $\Hil_0(\widetilde{\Lambda},\Spin L -N)$
by $U(\pi) \cdot \phi_A = \phi_{\pi A}$, where $\pi A(x,j) = A(x,\pi^{-1}(j))$.
By \eq{symmetrization ip}, we know that the range of $\widetilde{P}(\widetilde{\Lambda},N)$
is a trivial representation of $\mathfrak{S}_{2\Spin}$.
We define $\mathbb{P}_0(L,2\Spin,N)$ to be the set of all sequences $\mu = (\mu_1,\dots,\mu_{2\Spin})$
such that 
$$
L \geq \mu_1 \geq \mu_2 \geq \dots \geq \mu_{2\Spin} \geq 0
$$ 
and $\mu_1 + \dots + \mu_{2\Spin} = N$.
These are restricted partitions, but allowing parts equal to zero.
(We mention this fact for consistency. The interested reader can consult \cite{And},
\cite{GR}, or \cite{Sta} for more information about partitions.)
Then the range of $\widetilde{P}(\widetilde{\Lambda},N)$ is spanned by the linearly independent vectors
$$
\psi_\mu = \frac{1}{(2\Spin)!} \sum_{\pi \in \mathfrak{S}_{2\Spin}} 
  U(\pi) \Psi(\widetilde{\Lambda},\Spin L - \mu)\, ,\qquad
\mu \in \mathbb{P}_0(L,2\Spin,N)\, .
$$
We note that we can define an action of $\mathfrak{S}_{2\Spin}$ on $\Ir^{2\Spin}$ in the obvious way,
with the outcome that
$$
\Psi(\widetilde{\Lambda},\Spin L - \pi \boldsymbol{n}) = U(\pi) \Psi(\widetilde{\Lambda},\Spin L - \boldsymbol{n})\, .
$$

The orthogonal basis $\{\psi_\mu : \mu \in \mathbb{P}_0(L,2\Spin,N)\}$ is not yet orthonormal.
We observe that
\begin{align*}
\|\psi_\mu\|^2 &= ((2\Spin)!)^{-2} \|\Psi(\widetilde{\Lambda},\Spin L - \mu)\|^2 \\
&\qquad  \times \#(\textrm{orbit of } \mu) \times \#(\textrm{stabilizer of }\mu)^2\, .
\end{align*}
Since $\#(\textrm{stabilizer}) \times \#(\textrm{orbit}) = \# \mathfrak{S}_{2\Spin} = (2\Spin)!$,
we have
\begin{equation}
\begin{split}
\|\psi_\mu\|^2 &= 
  \#(\textrm{orbit of } \mu)^{-1} \|\Psi(\widetilde{\Lambda},\Spin L - \mu)\|^2 \\
  &= \binom{2\Spin}{n_0(\mu),n_1(\mu),n_2(\mu),\dots,n_L(\mu)}^{-1}
  \sum_{\substack{\boldsymbol{c} \in [0,2\Spin]^L \\ \sum_x c(x) = N}}
  q^{2 \boldsymbol{x} \cdot \boldsymbol{c}} M_{\boldsymbol{n},\boldsymbol{c}} 
\end{split}
\end{equation}
where $n_k(\mu)$ is the number of parts of $\mu$ equal to $k$.

By \eq{symmetrization ip}, and the fact that the inner product is invariant under
permutations of $\boldsymbol{m}$ and $\boldsymbol{n}$, we obtain
$$
\ip{\psi_\mu}{P_\Lambda \psi_\nu} 
  = \sum_{\substack{\boldsymbol{c} \in [0,2\Spin]^L \\ \sum_x c(x) = N}} M_{\boldsymbol{m},\boldsymbol{c}}
  M_{\boldsymbol{n},\boldsymbol{c}} q^{2\boldsymbol{x}\cdot \boldsymbol{c}} \prod_{x=1}^L \binom{2\Spin}{c(x)}\, .
$$
for all $\mu,\nu \in \mathbb{P}_0(L,2\Spin,N)$.
Therefore $\widetilde{P}(\widetilde{\Lambda},N)$ is represented, on its range, by a matrix
$\mathcal{P} = (\mathcal{P}(\mu,\nu) : \mu,\nu \in \mathbb{P}_0(L,2\Spin,N))$, where
\begin{equation}
\label{projected symmetrized matrix}
\begin{split}
\mathcal{P}(\mu,\nu) 
  &= \binom{2\Spin}{n_0(\mu),\dots,n_L(\mu)}^{1/2} 
  \binom{2\Spin}{n_0(\nu),\dots,n_L(\nu)}^{1/2} \\
  &\qquad \times \left(
  \sum_{\substack{\boldsymbol{c} \in [0,2\Spin]^L \\ \sum_x c(x) = N}} M_{\boldsymbol{m},\boldsymbol{c}}
  M_{\boldsymbol{n},\boldsymbol{c}} q^{2\boldsymbol{x}\cdot \boldsymbol{c}} \prod_{x=1}^L \binom{2\Spin}{c(x)}\right)\\
  &\qquad \Big/ \left(\sum_{\substack{\boldsymbol{c} \in [0,2\Spin]^L \\ \sum_x c(x) = N}}
  q^{2 \boldsymbol{x} \cdot \boldsymbol{c}} M_{\boldsymbol{n},\boldsymbol{c}}\right)\, .
\end{split}
\end{equation}
This is quite a complicated looking formula.
There are two nice features about it.
First, the powers of $q$ grow quite rapidly.
Second,  the matrix $(M_{\mu,\nu'} : \mu,\nu \in \mathbb{P}_0(L,2\Spin,N))$ is upper triangular 
with respect to dominance order,
where $\nu'$ is the transpose of the partition $\nu$.
(Dominance order is the natural partial order on partitions.)
These two facts insure that the matrix components of $\mathcal{P}$ decay exponentially
with the distance from the diagonal.
(More details and an equivalent expression are found in \cite{St2}.)

We have not attempted a rigorous analysis of $\mathcal{P}$, 
but we have obtained
very convincing numerical data, by simply numerically diagonalizing the matrix.
The main qualitative feature of the lower bound for the spectral gap is that it is not always
maximized at the Ising limit.
In particular, if $\Spin>3/2$ and the number of down spins $N$ satisfies
$N = 2 \Spin \floor{|\Lambda|/2}$, then the local
maximum for the lower bound of the spectral gap occurs somewhere other than the Ising limit.
The Ising limit is a classical model, whose energy levels can be calculated explicitly, and doing so
it is easy to see that the lower bound for the spectral gap is actually equal to the
true spectral gap at the Ising limit.
So the true spectral gap of the XXZ spin chain with $\Spin>3/2$ 
has a local maximum somewhere other than the Ising limit 	in finite volumes
as long as 
the number of down spins satisfies $N \approx \Spin |\Lambda|$ and $N \equiv 0 (\textrm{mod } 2\Spin)$.
One can take $N \approx \Spin |\Lambda|$ and $N \equiv 0 (\textrm{mod } 2\Spin)$ because
of the approximate periodicity of the spectral gap in $N$.
This same result is also obtained by Ising perturbation series, where we show that the curve
for the spectral gap is concave up at the Ising limit.
Also, the asymptotic analysis of Section 7 verifies the qualitative picture of the spectral
gap when $\Spin \gg 1$.

We now give an alternative description of $\mathcal{P}$ in terms of representations.
Recall that the ground state space of the spin 1/2 XXZ model is the highest
dimensional irreducible representation of $\textrm{SU}_q(2)$.
On the other hand, the symmetric tensors form the highest dimensional
irreducible representation of $\textrm{SU}(2)$.
Consider the array $\widetilde{\Lambda} = [1,L] \times [1,2\Spin]$.
At each site put a two dimensional representation of $\textrm{SU}(2)$ and 
$\textrm{SU}_q(2)$.
Note that this is possible because the two dimensional representations of $\textrm{SU}(2)$
and $\textrm{SU}_q(2)$ coincide.
Now tensor all the representations in a single row, considering them as representations of $\textrm{SU}_q(2)$.
For each row, define an operator $R_j(q)$ which projects onto the highest dimensional irreducible representation
of $\textrm{SU}_q(2)$.
Next tensor all the representations in a single column, considering them as representations of $\textrm{SU}(2)$.
Define $C_x$ to project onto the highest dimensional irreducible representation.
Then the operator $\sum_N \widetilde{P}(\widetilde{\Lambda},N)$ is identical to
$\prod_{x=1}^{L} C_x \prod_{j=1}^{2\Spin} R_j(q)$, modulo null spaces.
If one turns the procedure around, first projecting on columns then on rows, one almost
(but not quite) recovers the original problem of the spin $\Spin$ XXZ chain.


\subsection{Perturbation Series about Ising Limit}

We now perform a perturbation analysis for $\gamma([1,L],\Spin,M,\Delta^{-1})$
about the point $\Delta^{-1}=0$, i.e.\ the Ising limit.
We write
\begin{gather*}
H(\Delta^{-1}) = H^{(0)} + \Delta^{-1} H^{(1)} \\
H^{(0)} = \sum_{x=1}^{L-1} (\Spin + S_x^3)(\Spin - S_{x+1}^3) \\
H^{(1)} = \sum_{x=1}^{L-1} \left(-\frac{1}{2} S_x^+ S_{x+1}^- 
  - \frac{1}{2} S_x^- S_{x+1}^+\right)\, .
\end{gather*}
We have left out of $H^{(1)}$ the first order corrections to the boundary terms.
However since all our vectors are local perturbations of an Ising kink, the first order
corrections to the boundary terms will act as a multiple of the identity.
We will include these trivial corrections after we perform the perturbation theory
with $H^{(1)}$ as above.
We note that $H(\Delta^{-1})$ is unitarily equivalent to $H(-\Delta^{-1})$,
where the unitary transformation is
$$
U = \exp\left(2\pi i \sum_{j=1}^{\ceil{L/2}} S_{2j-1}^3\right)\, .
$$
This proves that the point $\Delta^{-1}=0$ is always either a local maximum or a local
minimum of $\gamma([1,L],\Spin,M,\Delta^{-1})$.
If $\gamma([1,L],\Spin,M,\Delta^{-1})$ is differentiable near $\Delta^{-1}=0$,
the first derivative is zero, and we proceed to second order perturbation theory.
The reason $\gamma([1,L],\Spin,M,\Delta^{-1})$ may not be differentiable near
$\Delta^{-1}=0$ is that the first excited state may be infinitely degenerate in the Ising limit.
This is the case for spin $1/2$ and for $\Spin=1$ when $M$ is odd.
For $\Spin=1/2$ and any $x \in \Ir$ there is an Ising ground state 
$$
\Psi_{0}(x) = \left(\bigotimes_{y \leq x} \ket{-1/2}_y\right) 
\otimes \left(\bigotimes_{y \geq x+1} \ket{1/2}_y\right)\, . 
$$
It is easy to see that there are infinite families of first excitations, for example
$\prod_{j=1}^L S_{x+1-j}^+ S_{y+j-1}^- \Psi_0(x)$ for any $L\geq 1$ and $y \geq x+2-L$.
For $\Spin=1$ and $M$ odd, the ground state is 
$$
\Psi_0(x) = \left(\bigotimes_{y<x} \ket{-1}_y\right) 
  \otimes \ket{0}_x \otimes \left(\bigotimes_{y>x} \ket{1}_y\right)\, ,
$$
and there are two classes of excitations, each infinitely degenerate:
$S_y^+ S_x^- \Psi_0(x)$ for any $y<x$; and 
$S_x^+ S_y^- \Psi_0(x)$ for any $y>x$.
These are the only cases where the first excitations are infinitely degenerate.
The only cases where there is a finite degeneracy for the first excited state
are $\Spin=2,3,4,\dots$ and $M$ congruent to $\Spin$ modulo $2\Spin$.
Then the ground state is 
$$
\Psi_0(x) = \left(\bigotimes_{y<x} \ket{-\Spin}_y\right) 
  \otimes \ket{0}_x \otimes \left(\bigotimes_{y>x} \ket{\Spin}_y\right)\, ,
$$
and the two first excitations are $S_{x-1}^+ S_x^- \Psi_0(x)$ and $S_x^+ S_{x+1}^- \Psi_0(x)$.
The other most interesting case, which has unique first excitations are $J\geq 1$ and
$M$ divisible by $2\Spin$.
Then
$$
\Psi_0(x) = \left(\bigotimes_{y\leq x} \ket{-\Spin}_y\right) 
  \otimes \left(\bigotimes_{y>x} \ket{\Spin}_y\right)\, ,
$$
and the unique first excitation is $S_x^+ S_{x+1}^- \Psi_0(x)$.
There are ground states with infinitely degenerate second excitations, and so on, but this
does not interest us.

We now consider the results of second order perturbation theory,
assuming the first order excitation is non-degenerate.
The kink ground states of the Ising model are all of the form
$$
\Psi_0(x,n)  = \left(\bigotimes_{y< x} \ket{-\Spin}_y\right) \otimes \ket{-\Spin+n}_x
  \left(\bigotimes_{y> x} \ket{\Spin}_y\right)\, ,
$$
where one can assume that $0 \leq n \leq \floor{\Spin}$.
The first excited state is then
$$
\Psi_1(x,n) = \left(\bigotimes_{y< x} \ket{-\Spin}_y\right) \otimes \ket{-\Spin+n+1}_x \otimes \ket{\Spin-1}_{x+1}
  \left(\bigotimes_{y> x+1} \ket{\Spin}_y\right)\, ,
$$
which has energy $E^{(0)} = n+1$.
We now expand to determine the corrections for small but nonzero $\Delta^{-1}$.
In particular we write $E(\Delta^{-1}) = E^{(0)} + \Delta^{-1} E^{(1)} + \dots$.
The perturbation series is standard, so we omit details.
The results are that $E^{(1)} = 0$ and 
$$
E^{(2)} = -\frac{1}{2} 
\left(2J(J-1) + n - \frac{2(J+1)(2J-1)}{n+3} + \frac{4 J^2}{2J-n-1}\right)\, .
$$
This is not an accurate description of the kink Hamiltonian because we have not included the
correct boundary fields.
To fix this situation we must add $\Spin (\sqrt{1 - \Delta^{-2}} - 1) (S_1^3 - S_L^3)$.
It is obvious that for a long enough spin chain, and excitations which are localized at the interface,
the extra boundary fields act just as $-2 \Spin^2 (\sqrt{1 - \Delta^{-2}} - 1)$ times the identity.
Note that this is $\Delta^{-2} J^2 + o(\Delta^{-2})$.
So for $0 \leq M < \Spin$, 
\begin{equation}
\begin{split}
&\left.\frac{d^2}{d (\Delta^{-1})^2}\right|_{\Delta^{-1}=0} \gamma([1,L],\Spin,M,\Delta^{-1}) \\
&\qquad\qquad  = \Spin - \frac{M}{2} + \frac{(\Spin+1)(2\Spin-1)}{M+3} - \frac{2 \Spin^2}{2\Spin - M - 1}\, .
\end{split}
\end{equation}
The finitely degenerate case is $\Spin \in \Ir$ and $M \equiv \Spin$ mod $(2\Spin)$, as mentioned before.
Then the first excitation of the Ising ground state is doubly degenerate, and we perform degenerate
perturbation theory.
As soon as $\Delta^{-1}>0$, the degeneracy lifts and there are two branches.
It is easily verified that the curvature of both branches is negative, but we
are only concerned with the lowest branch which gives
\begin{equation}
\left.\frac{d^2}{d (\Delta^{-1})^2}\right|_{\Delta^{-1}=0} \gamma([1,L],\Spin,\Spin,\Delta^{-1})
  = -8 - \frac{3}{\Spin-1} - \frac{\Spin}{2} + \frac{14}{\Spin+3}\, .
\end{equation}
We list some values for the curvature of $\gamma$ in Table \ref{tab:curvature}.
In the Ising limit, the minimum gap occurs for $n=0$.
One can see from this table that for $M=0$ and $\Spin>1$, the gap is concave
up at $\Delta^{-1}=0$.
This is the basis for Conjecture \ref{Ising conjecture}.
\begin{table}[t]\label{tab:curvature}\begin{center}
\begin{tabular}{c||c|c|c|c}
 &$n=0$&1&2&3\\ \hline\hline
$J=1/2$ & $-\infty$ &&&\\ \hline
1 & $-$1/3 & $-\infty$ &&\\ \hline
3/2 & 11/12 & $-$9/4&&\\ \hline
2 & 7/3 & $-$1/4 & $-$46/5 & \\ \hline
5/2 & 97/24 & 4/3 & $-$39/20 & \\ \hline
3 & 91/15 & 3 & 0 & $-$26/3
\end{tabular}
\end{center}
\caption{Some values of the curvature of the gap in the Ising limit.
For $\Spin=1/2$, $n=0$,and $\Spin=1$, $n=1$, the excited state
is infinitely degenerate, and the curvature is infinite as well.}\end{table}


\section{Boson Model}
We now give a heuristic derivation of the free Bose gas model for the XXZ
spin system in the limit $\Spin \to \infty$.
This is an approximation to the full XXZ Hamiltonian 
$H^\Spin_\Lambda$ on a finite chain $\Lambda = [1,L]$.
Our approach is similar to that of \cite{HP}, although we would suggest to the
reader to look at \cite{Dys1,Dys2} and \cite{CS} instead.
We are interested in the classical limit, $\Spin \to \infty$.
Lieb, \cite{L4}, proved that for the Heisenberg model
one obtains the classical partition
function as a scaled limit of quantum partition functions.
Lieb's method used coherent states to obtain rigorous upper and lower bounds
on the partition function.
In \cite{CS}, the same result was derived without coherent states,
and then it was shown that with a sufficiently large external magnetic field
the large $\Spin$ limit of the XXX model can be viewed as a free Bose gas,
which verified predictions of Dyson in \cite{Dys1,Dys2}.
A more recent proof of the Bose gas limit has been obtained by Michoel and Verbeure
\cite{MV}, where the spin $\Spin$ operator is viewed as a sum of $2\Spin$ spin $1/2$ operators,
(using a spin ladder), and then a noncommutative central limit theorem is applied.

The physical requirement of a large external field to obtain the Bose gas limit in the isotropic
case is easy to understand.
All the approximations (including our own) rely on a spin wave description of the elementary
excitations.
In order for this to be valid, the ground state must be very nearly saturated,
i.e.\ the ground state should satisfy $|\langle{S^3_\alpha}\rangle| - \Spin \ll \Spin$.
To accomplish this for the isotropic Heisenberg model, one must place a rather large
external magnetic field.
This is an important difference between the isotropic and anisotropic ferromagnets.
The XXZ model with $\Delta>1$ possesses kink ground states, which the isotropic model does not.
For the kink ground states, there is a quantum interface separating
two regions, which we can assume to be located at $\alpha = 1/2$.
For sites $\alpha \leq 0$ one has $\langle{S^3_\alpha}\rangle \leq -\Spin + C q^{-\alpha}$
and for $\alpha \geq 1$, $\langle{S^3_\alpha}\rangle \geq +\Spin - C q^{\alpha}$.
Thus the spin is saturated well away from the interface, with exponentially
small corrections.
The XXZ model exhibits saturation with just a boundary field, 
and the boundary field vanishes in the thermodynamic limit.
Moreover the boundary field is known to give the correct ground states (cf \cite{KN1}).
So the boson picture is quite natural for the XXZ model.

The energy--momentum dispersion relation is different for the XXZ model than for the isotropic
model, as one would expect.
The most important difference is that the lowest energy spin wave is not actually localized
in momentum space, but in position space.
It is localized at the interface, instead of being spread out
uniformly over a large region.
(There is one other spin wave with lower energy, in fact zero energy.
But this is the spin wave which simply moves one ground state to the other, owing to the fact that all
ground states in all sectors of total $S^3$ have equal energy.
We remove this boson by restricting to a single sector.)
Moreover, there is a spectral gap between the lowest spin wave, and the others.
The next independent spin wave boson does have a well-defined momentum,
and from there on the usual picture of spin waves prevails.
These are the results for one dimension, but the Bose gas model also holds for excitations
of the $(1,\dots,1)$ interface ground states in dimensions $d \geq 2$.
In dimensions higher than one, the low lying spectrum is more complex, having a continuous band 
of interface excitations at the bottom, as proved in \cite{BCNS,BCNS2}.
An interesting recent result by Caputo and Martinelli \cite{CapMart}
gives a rigorous lower bound for the spectral gap in a large but finite system $\Lambda$, whose power
law is $|\Lambda|^{-2/d}$ in agreement with \cite{BCNS}.
This gives strong evidence that the only excitations beneath a certain energy are
interface excitations.
The Bose gas approximation implies more, that for large $\Spin$ the interface excitations
are separated from all other spin wave excitations by a spectral gap of order $\Spin$.
The $d \geq 2$ results will be elucidated in a forthcoming paper
\cite{NS2}, as will be the rigorous proof of the spin wave
Boson model for the XXZ model.
For now we provide a heuristic argument.

We begin by considering the simplest case, namely $\Lambda = [1,2]$.
It is convenient to work in the dual space to $\Hil(\Lambda,\Spin)$.
Namely, for $\psi : [-\Spin,\Spin]\times [-\Spin,\Spin] \to \Cx$, define 
$$
\ket{\psi} = \sum_{m_1,m_2 \in [-\Spin,\Spin]} \psi(m_1,m_2) \ket{m_1,m_2}\, .
$$
Then $H_{\Lambda}^{\Spin} \ket{\psi} = \ket{G_\Spin \psi}$ 
where $G_{\Spin}$ is an operator on $\Cx^{[-\Spin,\Spin]^2}$
{\small
\begin{align*}
G_{\Spin} \psi(m_1,m_2) 
  &= (J^2 - m_1 m_2 + A(\Delta) J (m_1 - m_2)) \psi(m_1,m_2) \\
&\qquad -\frac{1}{2\Delta} [J(J+1) - m_1(m_1+1)]^{1/2} \\
&\qquad\qquad [J(J+1) - m_2(m_2-1)]^{1/2} \psi(m_1+1,m_2-1) \\
&\qquad  -\frac{1}{2\Delta} [J(J+1) - m_1(m_1-1)]^{1/2} \\
&\qquad\qquad [J(J+1) - m_2(m_2+1)]^{1/2} \psi(m_1-1,m_2+1) 
\end{align*}
}
where $A(\Delta) = \sqrt{1 - \Delta^{-2}}$, and we define $\psi(m_1,m_2) = 0$ for any
$(m_1,m_2)$ not in $[-\Spin,\Spin]\times [-\Spin,\Spin]$.
All we have done is to explicitly write down the action of the spin matrices.
Next, for any real numbers $-1<\mu_i<1$, $i=1,2$, let us define a linear operator 
$T_{\mu_1,\mu_2,\Spin} : \mathcal{C}^\infty(\Rl^2) \to \Cx^{[-\Spin,\Spin]^2}$, where
$T_{\mu_1,\mu_2,\Spin} \Psi = \psi$, 
$$
\psi(m_1,m_2) = \Psi(J^{-1/2} (m_1 - \mu_1 J), J^{-1/2} (m_2 - \mu_2 J))\, .
$$
This operator has a very large null space.
But if for fixed $\mu_1,\mu_2$ one knows that $T_{\mu_1,\mu_2,\Spin} \Psi = 0$ for every
$\Spin$, then $\Psi$ must obviously also be zero.

There are many choices of operators $\scrH_{\Spin}$ on $\mathcal{C}^\infty(\Rl^2)$
which satisfy $G_{\Spin} T_{\mu_1,\mu_2,\Spin} = T_{\mu_1,\mu_2,\Spin} \scrH_{\Spin}$.
One particularly good choice is the following
{\small
\begin{equation}
\begin{split}
\label{ExplicitFormula}
\scrH_\Spin \Psi(x_1,x_2) 
  &= \Spin^2 \bigg(1 - (\mu_1  + x_1 \Spin^{-1/2})(\mu_2  + x_2 \Spin^{-1/2})  \\
  &\hspace{50pt}
  + A(\Delta) \left( (\mu_1-\mu_2) + (x_1 - x_2) \Spin^{-1/2}\right)\bigg) \Psi(x_1,x_2)\\
  &\hspace{-50pt} - \frac{1}{2 \Delta} \sum_{\varepsilon = \pm 1}
  \Bigg( \prod_{i=1,2} \left[\Spin(\Spin+1) - (\mu_i \Spin + x_i \Spin^{1/2})
  (\mu_i \Spin + x_i \Spin^{1/2} + (-1)^{i+1} \varepsilon)\right]^{1/2}\\
  & \Psi\left(x_1 + \varepsilon \Spin^{-1/2}, x_2 - \varepsilon \Spin^{-1/2}\right) \Bigg)\, ,
\end{split}
\end{equation}
}
which is the same as the definition of $G_\Spin$, but now allowing the operator to act
on smooth functions instead of discrete functions.
One can formally expand
$$
\scrH_{\Spin} = \Spin^2 \scrH^{(2)} + \Spin^{3/2} \scrH^{(3/2)} + \Spin \scrH^{(1)} + \dots
$$
considering the shift by $\pm \Spin^{-1/2}$ as $e^{\pm \Spin^{-1/2} \partial}$, and expanding 
in the small parameter $\Spin^{-1/2}$.
The resulting expressions for $\scrH^{(2)}$, $\scrH^{(3/2)}$ and $\scrH^{(1)}$ are as follows
\begin{gather}
\label{H2FirstForm}
\scrH^{(2)} = 1 - \mu_1 \mu_2 + A(\Delta) (\mu_1 - \mu_2) - 
  \Delta^{-1} [1 - \mu_1^2]^{1/2} [1 - \mu_2^2]^{1/2}\, ,\\
\begin{split}
\scrH^{(3/2)} = A(\Delta) (x_1 - x_2) - \mu_1 x_2 - \mu_2 x_1 & \\
  &\hspace{-75pt}
  + \Delta^{-1} \left(\frac{\sqrt{1 - \mu_2^2}}{\sqrt{1 - \mu_1^2}} \mu_1 x_1
  + \frac{\sqrt{1 - \mu_1^2}}{\sqrt{1 - \mu_2^2}} \mu_2 x_2\right)\, ,
\end{split} \\
\label{H1FirstForm}
\begin{split}
\scrH^{(1)} &= - x_1 x_2 - \frac{1}{2 \Delta} [1 - \mu_1^2]^{1/2} [1 - \mu_2^2]^{1/2}\\
  &\bigg((\partial_{x_1} - \partial_{x_2})^2 - \frac{x_1^2}{(1-\mu_1^2)^2}
  - \frac{x_2^2}{(1-\mu_2^2)^2}
  + \frac{2 \mu_1 \mu_2 x_1 x_2}{(1 - \mu_1^2)(1 - \mu_2^2)} \\
  &\hspace{200pt}
  + \frac{1}{1 - \mu_1^2} + \frac{1}{1 - \mu_2^2}\bigg)
\end{split}
\end{gather}

Let us now use the parameter $\eta = \log(1/q)$, which is related to $\Delta$ by
$$
\Delta^{-1} = \sech(\eta)\, ,\qquad A(\Delta) = \tanh(\eta)\, .
$$
Then \eq{H2FirstForm} shows that $\scrH^{(2)}$ is a multiplication operator,
multiplying by the non-negative constant
$$
\sech(\eta) \left( e^{\eta/2} \sqrt{(1+\mu_1)(1-\mu_2)} - e^{-\eta/2} \sqrt{(1+\mu_2)(1-\mu_1)}\right)^2\, .
$$
Therefore, $\scrH^{(2)} \Psi = 0$ iff
\begin{equation}
\label{mu condition 1}
\exists r\in \Rl \textrm{ s.t. } \forall \alpha \in \Lambda\, ,\
\mu_\alpha = \tanh(\eta (\alpha - r))\, .
\end{equation}
The number $r$ is determined by $\mu_1 + \mu_2$, implicitly.
Note that $\scrH^{(3/2)}$ is also a multiplication operator, but given \eq{mu condition 1}
we know that it vanishes identically, as well.
So the first non-vanishing term is $\scrH^{(1)}$.
We use \eq{mu condition 1} to rewrite \eq{H1FirstForm}
\begin{equation}
\label{H1SecondForm}
\begin{split}
\scrH^{(1)}
  &= \frac{1}{2 \Delta} \operatorname{sech}(\eta(1-r)) 
  \operatorname{sech}(\eta(2-r)) \\
  &\times \bigg(-\partial_{x_1}^2 + \cosh^4(\eta(1-r)) x_1^2
  - \partial_{x_2}^2 + \cosh^4(\eta(2-r)) x_2^2 \\
  &\quad + 2 \partial_{x_1} \partial_{x_2}
  - 2 \cosh^2(\eta(1-r)) \cosh^2(\eta(2-r)) x_1 x_2 \\
  &\quad - \left[\cosh^2(\eta(1-r)) + \cosh^2(\eta(2-r))\right]\bigg)\, .
\end{split}
\end{equation}

Now we notice the following: $\scrH^{(1)}$ is a second order differential operator
which is homogeneous in $\partial_{x_1}$, $\partial_{x_2}$, $x_1$ and $x_2$ except for a 
constant (the zero point energy). Therefore, $\scrH^{(1)}$ can be regarded
as the Hamiltonian for a two-mode Boson system with quadratic interaction.
Thus, for $\alpha = 1,2$, we define 
\begin{equation}
\label{site boson definition}
\begin{split}
\hat{a}_\alpha &= \frac{1}{\sqrt{2}} \left( \cosh(\eta(\alpha-r)) x_\alpha 
  + \operatorname{sech}(\eta(\alpha-r)) \partial_{x_\alpha} \right)\, ,\\
\hat{a}_\alpha^\dagger &= \frac{1}{\sqrt{2}} \left( \cosh(\eta(\alpha-r)) x_\alpha 
  - \operatorname{sech}(\eta(\alpha-r)) \partial_{x_\alpha} \right)\, ,
\end{split}
\end{equation}
which satisfy the Canonical Commutation Relations
\begin{equation}
[\hat{a}_\alpha,\hat{a}_\beta] = [\hat{a}_\alpha^\dagger,\hat{a}_\beta^\dagger] = 0\, ,\qquad
[\hat{a}_\alpha,\hat{a}_\beta^\dagger] = \delta_{\alpha,\beta}\, ,
\end{equation}
and also
\begin{equation}
\hat{a}_\alpha^\dagger \hat{a}_\alpha
  = \frac{1}{2} 
\left(\cosh^2(\eta(\alpha-r)) x_\alpha^2 - \operatorname{sech}^2(\eta(\alpha-r)) \partial_{x_\alpha}^2 
  - 1\right)\, .
\end{equation}
One sees that
$$
\hat{a}_n + \hat{a}_n^\dagger = \sqrt{2} \cosh(\eta(n-r)) x_n 
$$
and
$$
\hat{a}_n - \hat{a}_n^\dagger = \sqrt{2} \operatorname{sech}(\eta(n-r)) 
\partial_{x_n}\quad .
$$
These relations imply
$$
2 \cosh(\eta(1-r)) \cosh(\eta(2-r)) x_1 x_2
  = (\hat{a}_1 + \hat{a}_1^\dagger)(\hat{a}_2 + \hat{a}_2^\dagger)
$$
and 
$$
2 \operatorname{sech}(\eta(1-r)) \operatorname{sech}(\eta(2-r)) \partial_{x_1}
  \partial_{x_2}
  = (\hat{a}_1 - \hat{a}_1^\dagger)(\hat{a}_2 - \hat{a}_2^\dagger)\, ,
$$
respectively. Hence
\begin{equation}
\begin{split}
&2 \partial_{x_1} \partial_{x_2}
  - 2 \cosh^2(\eta(1-r)) \cosh^2(\eta(2-r)) x_1 x_2 \\
  &\hspace{50pt}
  = - 2 \cosh(\eta(1-r)) \cosh(\eta(2-r)) (\hat{a}_1^\dagger \hat{a}_2
  + \hat{a}_2^\dagger \hat{a}_1)\, .
\end{split}
\end{equation}
All of these algebraic manipulations allow us to rewrite \eq{H1SecondForm} as
\begin{equation}
\label{H1FinalForm:2site}
\scrH^{(1)}
  = \frac{1}{\Delta} \bigg[ \frac{\cosh(\eta(1-r))}{\cosh(\eta(2-r))}
  \hat{a}_1^\dagger \hat{a}_1 + \frac{\cosh(\eta(2-r))}{\cosh(\eta(1-r))}
  \hat{a}_2^\dagger \hat{a}_2 
  - \hat{a}_1^\dagger \hat{a}_2 - \hat{a}_2^\dagger \hat{a}_1 \bigg]\, .
\end{equation}
One can verify that one of the two eigenmodes of this system has zero
energy, while the other is positive.
The zero mode is a direct consequence of the infinitely many ground states,
corresponding to different values of $M$, the third component of the total
spin.
If one considers the canonical picture, restricting the total magnetization
to have a fixed quantity, then that boson disappears.

The above results extend directly to a chain of arbitrary length.
One has $\mu_1,\dots,\mu_L$ satisfying \eq{mu condition 1}.
This is required for $\scrH^{(2)}$ to vanish, and sufficient for $\scrH^{(3/2)}$ 
to vanish.
The definition of the single site Bosons is just as in \eq{site boson definition},
for each $\alpha \in \Lambda = [1,L]$, and
the definition of $\scrH^{(1)}$ becomes
\begin{equation}
\scrH^{(1)}_\Lambda = \sum_{\alpha,\beta \in \Lambda} \mathcal{J}_{\alpha,\beta} 
\hat{a}_\alpha^\dagger \hat{a}_\beta\, ,
\end{equation}
where
$\mathcal{J}_{\alpha,\beta} = 0$ if $|\alpha-\beta|>1$,
$\mathcal{J}_{\alpha,\beta} = - \Delta^{-1}$ if $|\alpha-\beta|=1$, and
\begin{align*}
\mathcal{J}_{1,1} &=  \Delta^{-1} \frac{\cosh(\eta(1-r))}{\cosh(\eta(2-r))} \\
\mathcal{J}_{L,L} &=  \Delta^{-1} \frac{\cosh(\eta(L-r))}{\cosh(\eta(L-1-r))} \\
\mathcal{J}_{\alpha,\alpha} &= \Delta^{-1} \Big(\frac{\cosh(\eta(\alpha-r))}{\cosh(\eta(\alpha+1-r))} \\
&\qquad + \frac{\cosh(\eta(\alpha-r))}{\cosh(\eta(\alpha-1-r))} \Big)\, ,\qquad
\textrm{ for } 1<\alpha<L\, .
\end{align*}
The single site Bosons are coupled,
but the coupling matrix can be diagonalized. We obtain new Bose operators
$\hat{b}_n$, in terms of which the Hamiltonian is diagonal quadratic:
$$
\scrH^{(1)} = \sum_{n=0}^{L-1} \lambda_n \hat{b}^\dagger_n \hat{b}_n\, ,
$$
where $\hat{b}_n = \sum_{\alpha} v^{(n)}_\alpha \hat{a}_\alpha$ with
$v^{(n)}$ the eigenvector of $\mathcal{J}$ corresponding to the eigenvalue 
$\lambda_n$. Again there is one zero-mode, $\lambda_0=0$,
and $\lambda_i >0$, for $1\leq i\geq L-1$. $\lambda_1$ remains isolated in the
limit $L\to\infty$, for $0<q<1$.

By considering only the leading order terms for the bottom of the spectrum, we
have exchanged a quantum many body Hamiltonian $H^{\Spin}_\Lambda$ 
to an $L$ body Hamiltonian $\mathcal{J}$.
For sufficiently high spin this drastic simplification
describes the physics at low energies very well.

\begin{figure}
\begin{center}
\resizebox{5truecm}{5truecm}{\includegraphics{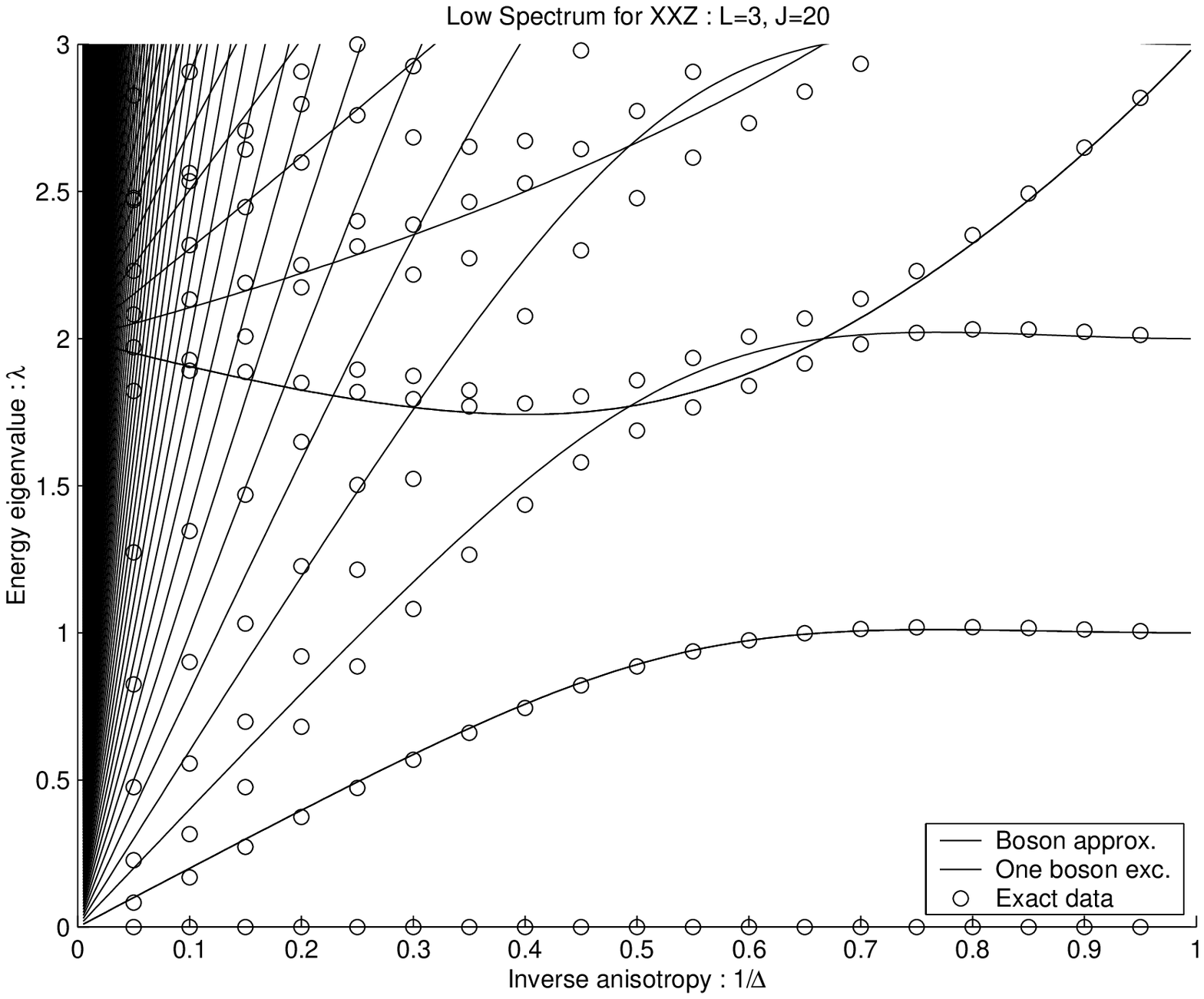}}\qquad
\resizebox{5truecm}{5truecm}{\includegraphics{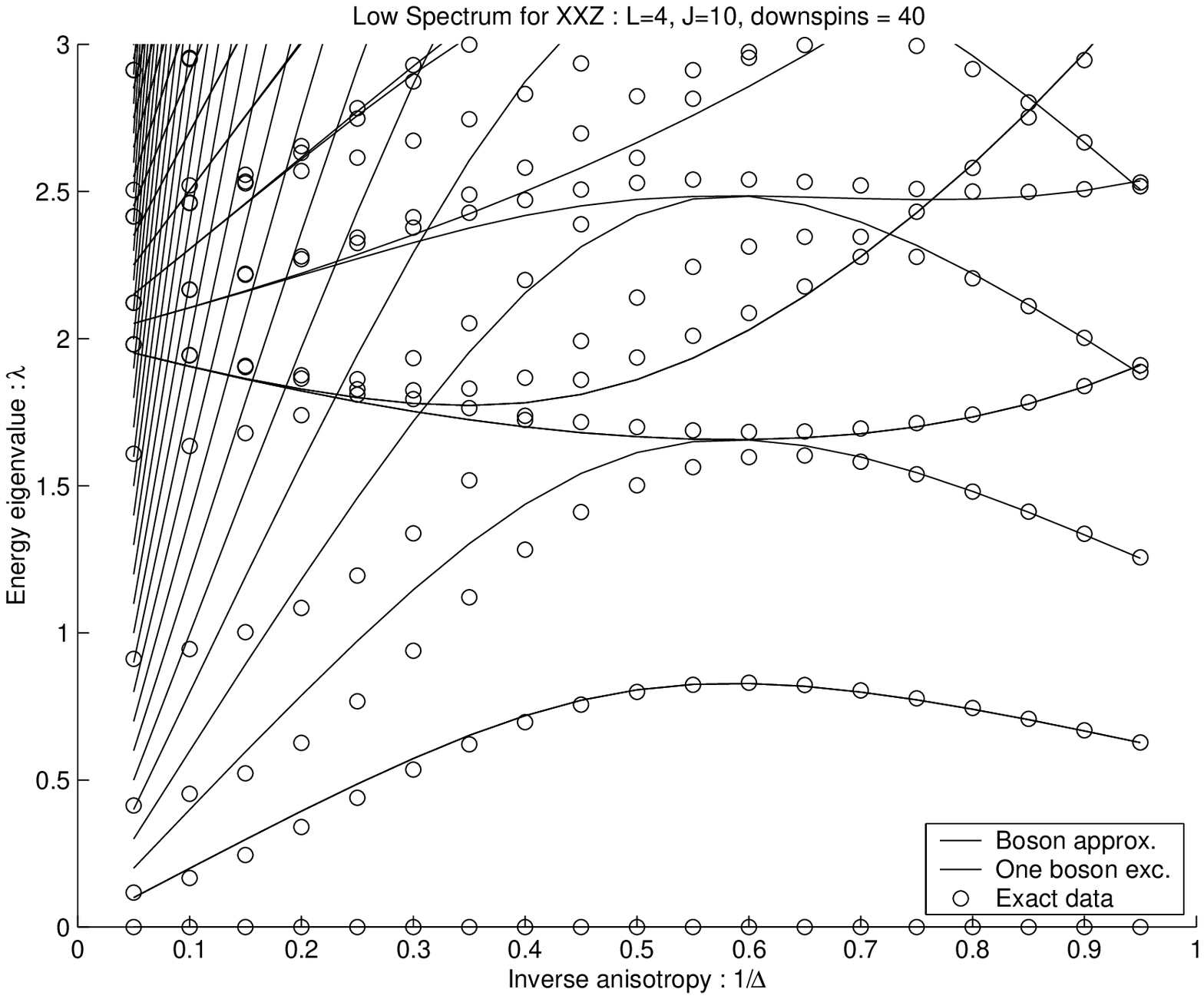}}\qquad
\end{center}
\caption{
\label{boson to actual2}
Solid lines are the predicted values of spectrum according to Boson
gas model, circles are actual values of the spectrum for full XXZ as obtained by Lanczos}
\end{figure}

\begin{figure}
\begin{center}
\resizebox{5truecm}{5truecm}{\includegraphics{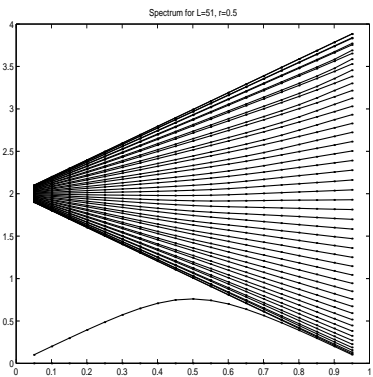}}\qquad
\end{center}
\caption{
\label{First J spec2}
The spectrum of the Boson coupling matrix versus anisotropy}
\end{figure}

\begin{figure}
\begin{center}
\resizebox{5truecm}{5truecm}{\includegraphics{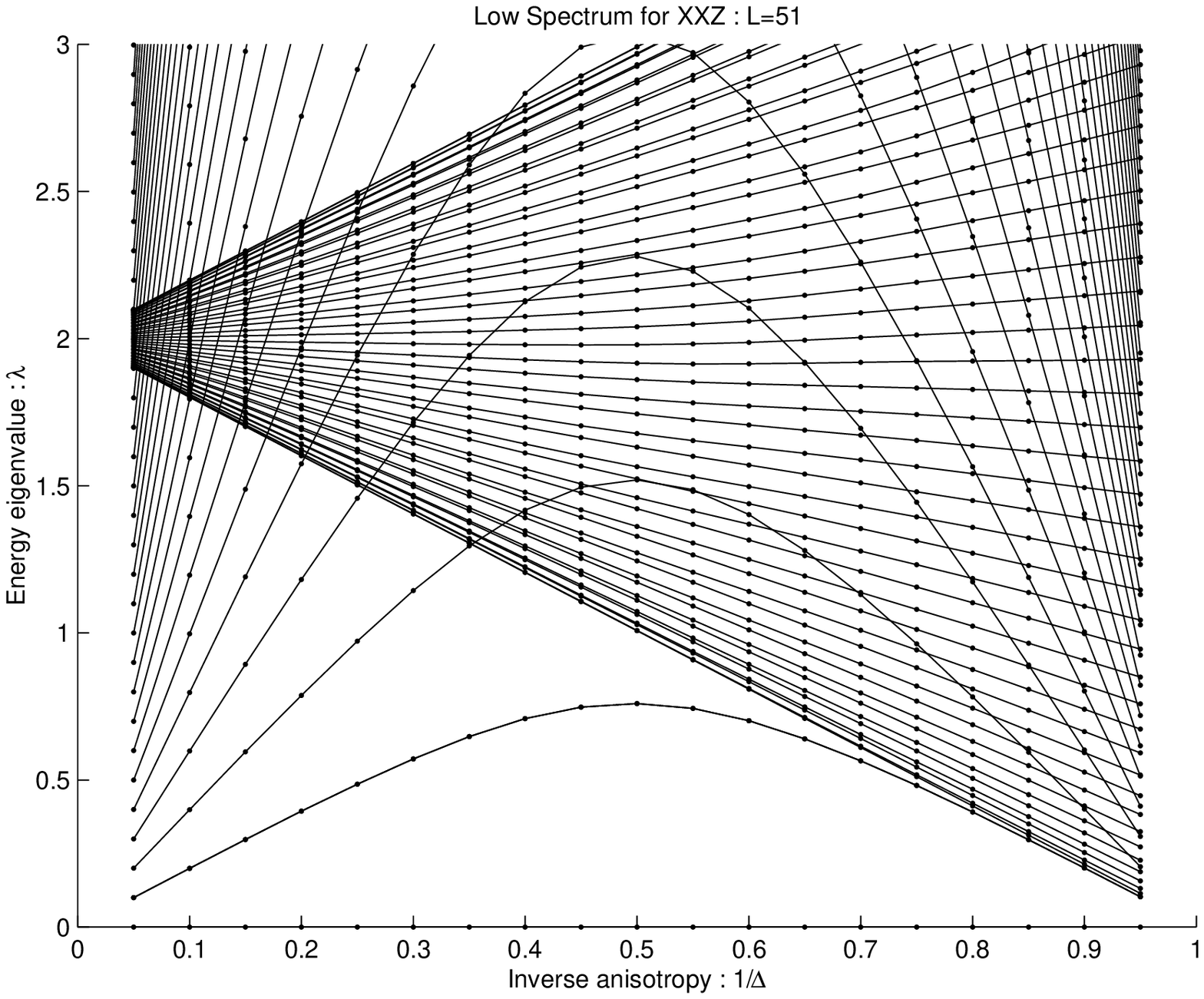}}\qquad
\end{center}
\caption{
\label{Second J spec2}
The spectrum of the Boson coupling matrix, with multiples
of the lowest boson energy included}
\end{figure}

In Figure \ref{boson to actual2} 
we compare our predictions to the spectrum of 
$H_\Lambda$ as obtained through numerical diagonalization. The comparison is
good, particularly along the first excited state and near the isotropic limit.
The reason that the Ising limit compares poorly is that the quantum
fluctuations have the effect of regularizing the eigenstates $\psi$, whereas
for the Ising model these states are definitely not smooth. However, for any
$q>0$, if $\Spin$ is made large enough, then we believe that these  asymptotics
eventually dominate. In Figure \ref{First J spec2}, we have plotted the spectrum
of $\mathcal{J}$ for fifty sites and $r=0$, which corresponds to the
magnetization for which one has a minimal gap. It is clear that there is only a
single isolated eigenvalue beneath  the branch of (what would in infinite
volumes be) continuous spectrum. Of course, to recover the spectrum of
$\scrH^{(1)}$ one must take all (nonnegative) integer valued  linear
combinations of these lines, since each line corresponds to the first excited
energy of an independent boson. In Figure \ref{Second J spec2},  we have plotted
several multiples of the eigenvalue line, to show how many eigenvalues lie
beneath the continuous spectrum. There is also interesting behavior for other
values of $r$. In Figure \ref{sgpic} we have plotted the spectral gap of
$\mathcal{J}$ as a function of $r$ and $\Delta^{-1}$ Finally we mention that
this analysis can be done for any dimension, not just one. Thus one may obtain,
at least heuristic, information about the low spectrum of  quantum spin systems
in higher dimensions by analyzing $\mathcal{J}_\Lambda$, which is the first
order quantum correction to the classical ground states for large but finite
$\Spin$. The higher-dimensional case, where the low-lying spectrum
is known to exhibit additional structure \cite{BCNS},
is the subject of a separate paper \cite{NS2}


\section*{Acknowledgements}

B.N. was partially supported by the National Science Foundation 
under Grant No. DMS0070774.
S.S. is a National Science Foundation Postdoctoral Fellow.

\nocite{Mat}
\nocite{BCN}
\bibliographystyle{hamsplain}
\bibliography{xxz}

\end{document}